\documentclass[a4paper,fleqn]{cas-dc}
\usepackage[numbers]{natbib}
\usepackage{algorithm} 
\usepackage[noend]{algpseudocode} 
\usepackage[titles]{tocloft}
\usepackage{tikz}
\usepackage{listings, listings-rust}
\usepackage{xspace}
\usepackage{tcolorbox}
\usepackage{multirow}
\usepackage{adjustbox}
\usepackage[inline]{enumitem}
\usepackage{xspace}
\usepackage{tcolorbox}
\usepackage{subcaption}
\usepackage{xcolor}

\makeatletter

\newenvironment{btHighlight}[1][]
{\begingroup\tikzset{bt@Highlight@par/.style={#1}}\begin{lrbox}{\@tempboxa}}
{\end{lrbox}\bt@HL@box[bt@Highlight@par]{\@tempboxa}\endgroup}

\definecolor{commentgreen}{RGB}{176, 176, 176}
\definecolor{rowcolor}{cmyk}{0,0.87,0.68,0.32}
\definecolor{rowcolor2}{cmyk}{ 20, 0, 37, 34}

\definecolor{eminence}{RGB}{108,48,130}
\definecolor{weborange}{RGB}{255,165,0}
\definecolor{frenchplum}{RGB}{129,20,82}
\definecolor{darkgreen}{RGB}{10, 92, 10}

\definecolor{celadon}{rgb}{0.67, 0.88, 0.69}

\newcommand\btHL[1][]{%
  \begin{btHighlight}[#1]\bgroup\aftergroup\bt@HL@endenv%
}
\def\bt@HL@endenv{%
  \end{btHighlight}
  \egroup
}
\newcommand{\bt@HL@box}[2][]{%
  \tikz[#1]{%
    \pgfpathrectangle{\pgfpoint{1pt}{0pt}}{\pgfpoint{\wd #2}{\ht #2}}%
    \pgfusepath{use as bounding box}%
    \node[anchor=base west, fill=orange!30,outer sep=0pt,inner xsep=1pt, inner ysep=0pt, rounded corners=3pt, minimum height=\ht\strutbox+1pt,#1]{\raisebox{1pt}{\strut}\strut\usebox{#2}};
  }%
}
\makeatother

\newcommand{\tool}{{\sc Wasm-Mutate}\xspace}
\newcommand{\Wasm}{WebAssembly\xspace}
\newcommand{\wasm}{\Wasm}
\newcommand{\etal}{et.al.\xspace}
\newcommand{\ie}{i.e.,\xspace}

\newtheorem{metric}{Metric}

\newlistof{todo}{td}{List of TODOs}

\newenvironment{revision1}{}{}
\newcommand{\revision}[1]{{#1}}
\newcommand{\repourl}{\url{https://github.com/bytecodealliance/wasm-tools/tree/main/crates/wasm-mutate}}
\newcommand{\dataurl}{\url{https://github.com/ASSERT-KTH/tawasco}}

\newcommand*\step[1]{
\noindent\tikz[baseline=(char.base)]{
        \node[shape=circle,text=black,draw=black, fill=white,inner sep=1.2pt] (char) {#1};}}

\lstdefinelanguage{WAT}{
    otherkeywords={},
    morekeywords=[1]{i32,f32,i64,f64,funcref, nop},
    morekeywords=[2]{0},
    morekeywords=[3]{add,or,const,mul,eqz,shl,get,rem_s,rem_u,ne,tee,sub,set,store},
    morekeywords=[4]{},
    morekeywords=[5]{global, get_global, mut, set_global, export, import,loop, memory, data, get_local,if,element, block,module, function, set_local,call,br_if,end,br, else, all,call_indirect,local,global,module, func, param, result, type, table, memory, elem},
    morekeywords=[6]{=,;},
    morekeywords=[7]{(,),[,],.},
    sensitive=false,
    morecomment=[l]{;},
    morecomment=[s]{;}{;},
    morestring=[b]",
    keywordstyle=[1]\color{eminence}\bfseries,
    keywordstyle=[3]\color{frenchplum},
    keywordstyle=[5]\color{darkgreen}\bfseries,
    commentstyle=\color{commentgreen}
}
\lstdefinestyle{nccode}{
        numbers=none,
        firstnumber=2,
        stepnumber=1,
        numbersep=10pt,
        tabsize=4, 
        showspaces=false,
        breaklines=true, 
        showstringspaces=false,
    moredelim=**[is][{\btHL[fill=weborange!40]}]{`}{`},
    moredelim=**[is][{\btHL[fill=celadon!40]}]{!}{!}
}

\lstdefinestyle{WATStyle}{
  numbers=left,
  stepnumber=1,
  numbersep=10pt,
  tabsize=4,
  showspaces=false,
  showstringspaces=true,
}

 \newcommand\tikzmarkPROBE[4]{%
\tikz[remember picture, overlay]{%
        \pgfusepath{use as bounding box}%
        \node[right=#3 mm,text width=2mm,align =center,outer sep=0pt,inner xsep=1pt, inner ysep=0pt, rounded corners=3pt,anchor=north, minimum height=#4 mm, text depth = #4 mm] at (1 mm,2 mm) (#1) {};
  }
 }

 \lstdefinelanguage{ttt}{
    otherkeywords={},
    morekeywords=[1]{i32,f32,i64,f64,funcref,x,nop},
    morekeywords=[2]{0},
    morekeywords=[3]{container, add,const,mul,shl,get,rem_s,eqz,rem_u,ne,tee,sub,set,store,or},
    morekeywords=[4]{},
    morekeywords=[5]{global, get_global, mut, set_global, export, import,loop, memory, data, get_local,if, block,module, set_local,call,br_if,end,br, else, ret,all,call_indirect,local,global,module, func, param, result, type, table, (, ), memory, elem},
    morekeywords=[6]{=,;},
    morekeywords=[7]{[,],.},
    morekeywords=[8]{LHS, RHS, Cond},
    sensitive=false,
    morecomment=[l]{;},
    morecomment=[s]{;}{;},
    morestring=[b]",
    keywordstyle=[1]\color{eminence}\bfseries,
    keywordstyle=[3]\color{frenchplum},
    keywordstyle=[5]\color{darkgreen}\bfseries,
    keywordstyle=[7]\color{darkgreen}\bfseries,
    keywordstyle=[8]\color{frenchplum}\bfseries,
    keywordstyle=[2]\color{darkgreen}\bfseries,
    commentstyle=\color{commentgreen}
}
 \renewcommand{\texttt}[1]{\lstinline{#1}}

 \lstdefinelanguage{TRACE}{
    otherkeywords={},
    morekeywords=[1]{dword, qword,ptr},
    morekeywords=[2]{0},
    morekeywords=[3]{mov, value, size, Writ, Read, I},
    morekeywords=[4]{rdx, ecx, r14},
    morekeywords=[5]{global, get_global, mut, set_global, export, import,loop, memory, data, get_local,if,element, block,module, function, set_local,call,br_if,end,br, else, all,call_indirect,local,global,module, func, param, result, type, table, memory, elem},
    morekeywords=[6]{=,;},
    morekeywords=[7]{(,),[,],.},
    sensitive=false,
    morecomment=[l]{;},
    morecomment=[s]{;}{;},
    morestring=[b]",
    keywordstyle=[1]\color{eminence}\bfseries,
    keywordstyle=[3]\color{frenchplum},
    keywordstyle=[5]\color{darkgreen}\bfseries,
    commentstyle=\color{commentgreen}
}

\begin{document}\sloppy


\newcommand\mytitle{\tool: Fast and Effective Binary Diversification for WebAssembly}
\shorttitle{}
\shortauthors{Cabrera-Arteaga et~al.}

\title [mode = title]{\mytitle}

\author[kth-address]{Javier Cabrera-Arteaga}[orcid=0000-0001-9399-8647]\cormark[1]\ead{javierca@kth.se}
\author[fastly-address]{Nicholas Fitzgerald,}[orcid=0000-0002-0209-2805]\ead{nfitzgerald@fastly.com}
\author[kth-address]{Martin Monperrus}[orcid=0000-0003-3505-3383]\ead{monperrus@kth.se}
\author[kth-address]{Benoit Baudry}[orcid=0000-0002-4015-4640]\ead{baudry@kth.se}
\cortext[cor1]{Corresponding authors}

\address[kth-address]{KTH Royal Institute of Technology, Stockholm, Sweden}
\address[fastly-address]{Fastly Inc., San Francisco, USA }

\hyphenation{Web-Assembly}
\hyphenation{Java-Script}
\hyphenation{equi-valent}



\definecolor{myblue}{RGB}{0,0,255}
\begin{abstract}
    WebAssembly is the fourth officially endorsed Web language.
    It is recognized because of its efficiency and design, focused on security. 
    Yet, its swiftly expanding ecosystem lacks robust software diversification systems. 
    We introduce \tool, a diversification engine specifically designed for WebAssembly. 
    Our engine meets several essential criteria: 
    1) To quickly generate functionally identical, yet behaviorally diverse, WebAssembly variants, 
    2) To be universally applicable to any WebAssembly program, irrespective of the source programming language, and 
    3) Generated variants should counter side-channels. 
    By leveraging an e-graph data structure, \tool is implemented to meet both speed and efficacy.
    We evaluate \tool by conducting experiments on 404 programs, which include real-world applications.
    Our results highlight that \tool can produce tens of thousands of unique and efficient WebAssembly variants within minutes.
    Significantly, \tool can safeguard \wasm binaries against timing side-channel attacks, especially those of the Spectre type.
\end{abstract}




\maketitle


\section{Introduction}

\Wasm(Wasm) is the fourth official language of the web, complementing HTML, CSS and JavaScript as a fast, platform-independent bytecode  \cite{haas2017bringing, WebAssemblyCoreSpecification}. 
Since its introduction in 2015, it has seen rapid adoption, with support from all major web browsers. 
\wasm has also been adopted outside of browsers, e.g., platforms like Fastly and Cloudflare use Wasm as their core technology \cite{fastly}.
Recently, in addition to major ones like LLVM, more compilers and tools can output \wasm binaries \cite{hilbig2021empirical, javy, kmm}. 
With this prevalence, software protection techniques for \wasm are remarkably needed \cite{avengers}.

Software diversification is a well-known software protection technique \cite{cohen1993operating, 4197960, 595185}, consisting of producing numerous variants of an original program, each retaining equivalent functionality. 
Software diversification in \wasm has several application domains, such as optimization \cite{superwasm} and malware evasion \cite{CABRERAARTEAGA2023103296} research. 
It can also be used for fuzzing, an example of this was the discovery of a CVE in Fastly in 2021 \cite{CVE}, achieved through transforming \wasm binary with functionally equivalent code replacements.

To develop an effective \wasm diversification engine, several key requirements must be met. 
First, the engine should be language-agnostic, enabling diversification of any \Wasm code, regardless of the source programming language and compiler toolchain.
Second, it must have the capability to swiftly generate functionally equivalent variants of the original code. 
The speed at which this diversification occurs holds potential for real-time applications, including moving target defense \cite{MEWE}. 
The engine should also possess the ability to harden potential attacks by producing sufficiently distinct code variants.
This paper presents an original engine, \tool, that addresses all these requirements.

\tool is a tool that automatically transforms a \wasm binary program into variants that preserve the original functionality. 
The core of the diversification engine relies on an e-graph data structure \cite{10.1145/3434304}.
To the best of our knowledge, this work is the first to use an e-graph for software diversification in \Wasm. An e-graph offers one essential property for diversification:  every path through the e-graph represents a functionally equivalent variant of the input program \cite{10.1145/3434304, 10.1145/3385412.3386001}.  
A random e-graph traversal is virtually costless, supporting the generation of tens of thousands of equivalent variants from a single seed program in minutes \cite{10.1145/3547622}. 
Consequently, the choice of e-graphs is the key to build a diversification tool that is both effective and fast.



\begin{revision1}
    
    We evaluate \tool by examining its ability to generate \Wasm variants.
    We have based our empirical evaluation on an existing corpus from the diversification literature \cite{arteaga2020crow, hilbig2021empirical, Swivel}.
    Additionally, we measure the speed at which \tool can produce the first variant that demonstrates a trace different from the original.
    Significantly, we evaluate the performance impact on generated variants of real-world \Wasm programs \cite{hilbig2021empirical}.
    Our security assessment of \tool involves determining the extent to which diversification can safeguard against Spectre attacks. 
    We carry out this assessment using \wasm programs previously identified as susceptible to Spectre attacks \cite{Swivel}.
\end{revision1}

Our results demonstrate that \tool can generate thousands of variants in minutes. These variants have unique machine code after compilation with Cranelift\footnote{\url{https://cranelift.dev/}} (static diversity) and the variants exhibit different traces at runtime (dynamic diversity).
\revision{Moreover, we empirically demonstrate that, in the worst scenario, the performance impact on the generated variants is maintained within the same order of magnitude as the original program.}
Remarkably, our experiments provide evidence that the generated variants are hardened against Spectre attacks.
To sum up, the contributions of this work are:

\begin{itemize}
    \item The design and implementation of a \Wasm diversification engine, based on semantic-preserving binary rewriting rules.
    \item Empirical evidence of the diversity of variants created by \tool, both in terms of static binaries and execution traces.
    \revision{
        \item Empirical evidence of the performance impact of the variants created by \tool in terms of size and execution time.
    }
    \item A clearcut demonstration that \tool can protect \wasm binaries against timing side-channel attacks, specifically, Spectre.
    \item An open-source repository, where \tool is publicly available for future research \repourl.    
\end{itemize}

This paper is structured as follows. 
In \autoref{background}, we introduce WebAssembly, the concepts of semantic equivalence, and what we state as a rewriting rule.
In \autoref{tech}, we explain and detail the architecture and implementation of \tool.
We formulate our research questions in \autoref{eval}, answering them in \autoref{results}.
We discuss open challenges related to our research in \autoref{discussion}, to help future research projects on similar topics.
In \autoref{rw} we highlight works related to our research on software diversification.
We finalize with our conclusions \autoref{conc}.

\section{Background}
\label{background}

In this section, we define and formulate the foundation of this work: WebAssembly and its runtime structure.
We also enunciate the notions of rewriting rules in the context of our work.

\subsection{WebAssembly}

WebAssembly (Wasm) is a binary instruction set initially meant for the web. 
It was adopted as a standardized language by the W3C in 2017 \cite{haas2017bringing}. One of Wasm's primary advantages is that it defines its own Instruction Set Architecture (ISA), making it platform-independent. 
As a result, a Wasm binary can execute on virtually any platform, including web browsers and server-side environments. 
WebAssembly programs are compiled ahead-of-time from source languages such as C/C++, Rust, and Go, utilizing compilation pipelines like LLVM. 
A \Wasm binary packages a collection of sections.
Each one of these sections could be optional and might contain specific restrictions.
For example, some sections must follow a relative order concerning other sections.
The organization of the \Wasm binary into sections boosts the validation and compilation of the binary once on the host engines.

At runtime, WebAssembly programs operate on a virtual stack that holds primitive data types.
Such data is then operated by typed stack instructions.
A WebAssembly program also declares linear memory and globals, which are used to store, manipulate, and share data during program execution, e.g. to share data with the host engine of the WebAssembly binary.
In \autoref{example:cprogram}, we provide an example of a Rust program that contains a function declaration, a loop, a loop conditional, and a memory access. When the Rust code is compiled to WebAssembly, it produces the code shown in \autoref{example:wasmprogram}. 

{
  \lstset{
   style=nccode,
    language=Rust,
    basicstyle=\footnotesize\ttfamily,
    columns=fullflexible,
  postbreak=\mbox{\textcolor{red}{$\hookrightarrow$}\space},
    breaklines=true}
\begin{lstlisting}[label=example:cprogram,caption={A Rust program containing function declaration, loop, conditional and memory access.},captionpos=b]{Name}
fn main() {
    let mut arr = [1, 2, 3, 4, 5];
    // Variable assignment
    let mut sum = 0;
    // Loop and memory access
    for i in 0..arr.len() {
        sum += arr[i];
    }
    // Use of external function
    println!("Sum of array elements: {}", sum);
}
\end{lstlisting}   
}

\lstdefinestyle{watcode}{
  numbers=none,
  stepnumber=1,
  numbersep=10pt,
  tabsize=4,
  showspaces=false,
  breaklines=true, 
  showstringspaces=false,
  moredelim=**[is][{\btHL[fill=black!10]}]{`}{`},
  moredelim=**[is][{\btHL[fill=celadon!40]}]{!}{!}
}
{
\lstset{
  language=WAT,
  style=watcode,
  breaklines=true, 
  basicstyle=\footnotesize\ttfamily,
  numbersep=2.5pt,
  escapeinside={``},
  }
    \begin{lstlisting}[label=example:wasmprogram,caption={Simplified WebAssembly code for the program of \autoref{example:cprogram}.}, captionpos=b]{Name}
(module
  (@custom "producer" "llvm.." )
  (import "env" "println" (func $println (param i32)))
  (memory 1)
  (export "memory" (memory 0))
  (func $main
    (local $sum i32)
    (local $i i32)
    (local $arr_offset i32)
    ; Initialize sum to 0 ;
    i32.const 0
    local.set $sum
    ; Initialize arr_offset to point to start of the array in memory ;
    i32.const 0
    local.set $arr_offset
    ; Initialize the array in memory;
    i32.const 0
    i32.const 1
    i32.store
    ...
    i32.store
    ...
    loop
      local.get $i
      i32.const 5
      i32.lt_s
      if
        ; Load array[i] and add to sum ;
        local.get $arr_offset
        local.get $i
        ...
        ; Increment i ;
        local.get $i
        i32.const 1
        i32.add
        local.set $i
        br 0
      else
        ; End loop ;
        i32.const 0
      end
    end
    
    ; Call external function to print sum ;
    local.get $sum
    call $println
  )
  ; Start the main function ;
  (start $main)
  )
\end{lstlisting}
}

WebAssembly is designed with isolation as a primary consideration, usually referred as Software Fault Isolation (SFI). 
For instance, a WebAssembly binary cannot access the memory of other binaries or cannot interact directly with the browser's APIs, such as the DOM or the network. Instead, communication with these features is constrained to functions imported from the host engine, ensuring a secure and safe WebAssembly environment.
Moreover, control flow in WebAssembly is managed through explicit labels and well-defined blocks, which means that jumps in the program can only occur inside blocks, unlike regular assembly code \cite{10.1145/3062341.3062363}.
\revision{
Besides, function tables are constructed statically.
Overall, the "least privilege" principle is the fundamental security requirement. 
This principle is reflected in the lack of certain features that are common in other programming environments.
For example, reflection, a feature available in many high-level programming languages like Java and C\#, is not possible in \Wasm. 
}


The WebAssembly runtime structure is described in the WebAssembly specification and it includes 10 key elements: the Store, Stack, Locals, Module Instances, Function Instances, Table Instances, Memory Instances, Global Instances, Export Instances, and Import Instances. These components interact during the execution of a WebAssembly program, collectively defining the state of a program during its runtime.
Yet, three of the previous runtime components, the Stack, Memory, and Globals, are particularly significant in maintaining the state of a WebAssembly program during its execution. 
The Stack holds both values and control frames, with control frames handling block instructions, loops, and function calls. 
The Memory represents the linear memory of a WebAssembly program, consisting of a contiguous array of bytes.
The Globals save data that is globally accessible by any code inside the \Wasm program and that can be optionally accessed by the host engine.
In this paper, we highlight the aforementioned three components to define, compare, and validate the state of WebAssembly programs during their execution.

\subsection{Rewriting rules}
\label{rewriting}

Our definition of a rewriting rule draws from the one proposed by Sasnauskas et al. \cite{2017arXiv171104422S}, and integrates a predicate to specify the replacement condition.
Concretely, a rewriting rule is defined as a tuple, denoted as \texttt{(LHS, RHS, Cond)}. Here, \texttt{LHS} refers to the code segment slated for replacement, \texttt{RHS} is the proposed replacement, and \texttt{Cond} stipulates the conditions under which the replacement is acceptable.
Importantly, \texttt{LHS} and \texttt{RHS} are meant to be functionally equivalent.

We focus on rewriting rules that guarantee functional equivalence. 
Functional equivalence refers to the notion that two programs are considered equivalent if, for the same input of the same domain they produce the same output\cite{10.1145/2594291.2594334}. 

For example, the rewriting rule \texttt{(x,\ x\ i32.or\ x, \{\})} implies that the \texttt{LHS} 'x' is to be replaced by an idempotent bitwise \texttt{i32.or} operation with itself, absent any specific conditions.
Notice that, for this specific rule, the commutative property shared by \texttt{LHS} and \texttt{RHS}, symbolized as \texttt{(LHS, RHS) = (RHS, LHS)}.
Besides, the \texttt{Cond} element could be an arbitrary criterion. 
For instance, the condition for applying the aforementioned rewriting rule could be to ensure that the newly created binary file does not exceed a threshold binary size.

Based on our understanding, our research is one of the first to apply the concept of rewriting rules to WebAssembly.
This will expand the potential use cases of \tool. 
Beyond its role as a diversification tool, it can also be used as a standard tool for conducting program transformations in WebAssembly.

\section {Design of \tool}
\label{tech}

In this section, we present \tool, a tool to diversify
WebAssembly binaries and produce functionally equivalent variants.
\begin{figure*}[h!]
    \centering
    \includegraphics[width=0.9\linewidth]{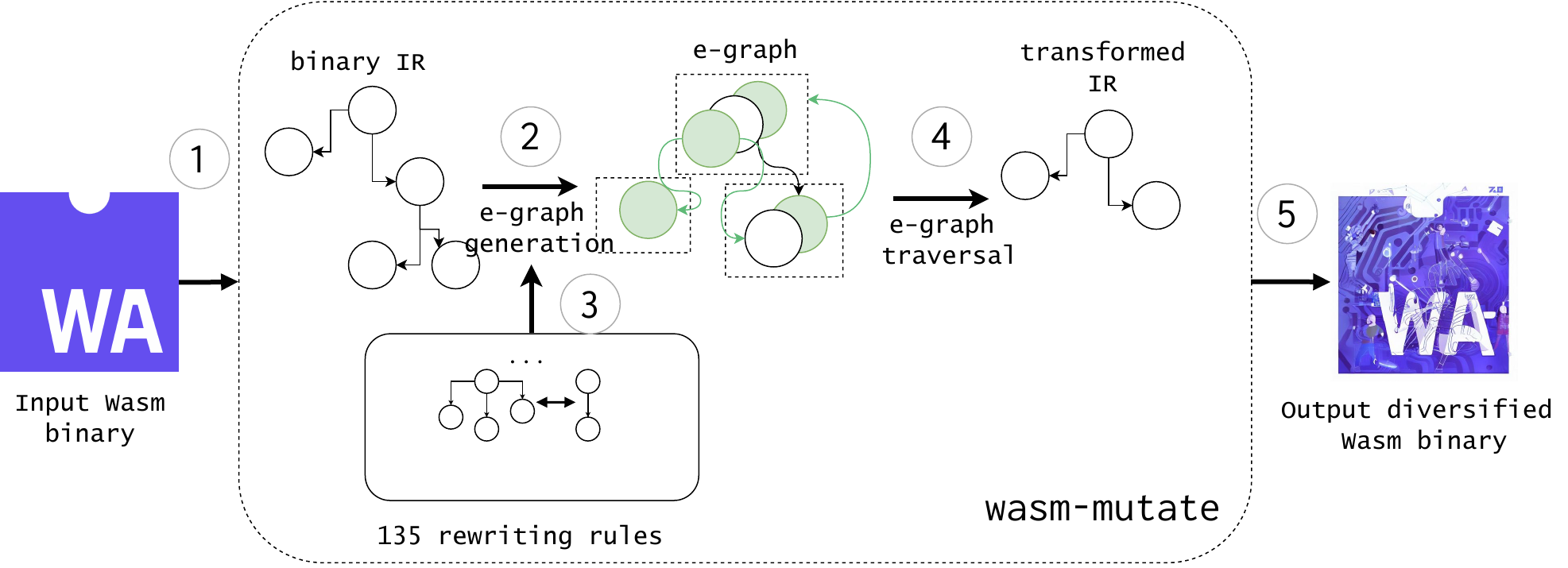}
    \caption{ \tool high-level architecture.  It generates semantically equivalent variants from a given WebAssembly binary input. 
    Its central approach involves synthesizing these variants by substituting parts of the original binary using rewriting rules, boosted by diversification space traversals using e-graphs(refer to \autoref{alg}).}
  \label{fig:wasm-mutate}
\end{figure*}

\subsection{Overview}
The primary objective of \tool is to perform diversification, i.e., generate functionally equivalent variants from a given WebAssembly binary input. 
\tool's central approach involves synthesizing these variants by substituting parts of the original binary using rewriting rules. 
It leverages a comprehensive set of rewriting rules, boosted by diversification space traversals using e-graphs.

In \autoref{fig:wasm-mutate} we illustrate the workflow of \tool: it starts with a WebAssembly binary as input \step{1}.
It parses the original binary \step{2}, turning the input program into appropriate abstractions, in particular, \tool builds the control flow graph and data flow graph. 
Using the defined rewriting rules, \tool builds an e-graph \step{3} for the original program.
An e-graph packages every possible equivalent code derivable from the given rewriting rules  \cite{10.1145/3434304, 10.1145/3385412.3386001}.
Thus, at this stage, \tool exploits a key property of e-graphs:
any path traversal through the e-graph results in a semantically equivalent code.
Then, the diversification process starts, with parts of the original program being randomly replaced by the result of traversing the e-graph \step{4}.
The outcome of \tool is a functionally equivalent variant of the original binary \step{5}.
\tool guarantees functionally equivalent variants because each rewriting rule is semantic preserving.


\subsection{WebAssembly Rewriting Rules}



In total, there are 135 possible rewriting rules implemented in \tool, those rules are grouped under several categories, called hereafter meta-rules.
For example, 125 rewriting rules are implemented as part of a peephole meta-rule.
In the following, we present 7 meta-rules.

\textbf{Add type:}
In WebAssembly, the type section wraps definitions of signatures for the binary functions.
\tool implements two rewriting rules, one of which is illustrated in the following. 

{\footnotesize
\lstdefinestyle{watcode}{
  numbers=none,
  stepnumber=1,
  numbersep=10pt,
  tabsize=4,
  showspaces=false,
  breaklines=true, 
  showstringspaces=false,
    moredelim=**[is][{\btHL[fill=weborange!40]}]{`}{`},
    moredelim=**[is][{\btHL[fill=celadon!40]}]{!}{!},
    moredelim=**[is][{\btHL[fill=frenchplum!40]}]{bb}{bb},
    moredelim=**[is][{\btHL[fill=eminence!40]}]{bc}{bc}
}

\lstset{
        language=ttt,
        style=watcode,
        basicstyle=\footnotesize\ttfamily,
        columns=fullflexible,
        breaklines=true}
\begin{lstlisting}[numbers=none]{Name}
LHS (module
  (type (;0;) (func (param i32) (result i64)))
        \end{lstlisting}  
\hrulefill

\begin{lstlisting}[numbers=none]{Name}
RHS (module
  (type (;0;) (func (param i32) (result i64)))
!+ (type (;0;) (func (param i64) (result i32 i64)))!
        \end{lstlisting}   
}

This transformation generates random function signatures with a random number of parameters and results count.
This rewriting rule does not affect the runtime behavior of the variant.
It also guarantees that the index of the already defined types is consistent after the addition of a new type. This is because WebAssembly programs cannot access or use a type definition during runtime, they are only used to validate the signature of a function during compilation and validation in the host engine.
From the security perspective, this transformation prevents static binary analysis. 
For example, to avoid malware detection based on a signature set \cite{CABRERAARTEAGA2023103296}.

\textbf{Add function:} The function and code sections of a WebAssembly binary contain function declarations and the code body of the declared functions, respectively.
To add a new function, \tool first creates a random type signature.
Then, the random function body is created.
The body of the function consists of returning the default value of the result type.
The following example illustrates this rewriting rule.

\lstdefinestyle{watcode}{
  numbers=none,
  stepnumber=1,
  numbersep=10pt,
  tabsize=4,
  showspaces=false,
  breaklines=true, 
  showstringspaces=false,
    moredelim=**[is][{\btHL[fill=black!10]}]{`}{`},
    moredelim=**[is][{\btHL[fill=celadon!40]}]{!}{!}
}

{
\captionsetup{width=\linewidth}
\noindent\begin{minipage}[b]{\linewidth}
    \lstset{
        language=ttt,
        style=watcode,
        basicstyle=\footnotesize\ttfamily,
        columns=fullflexible,
        breaklines=true}
        \begin{lstlisting}[]{Name}
 LHS (module
    (type (;0;) (func (param i32 f32) (result i64)))
        \end{lstlisting}
   \end{minipage}
   
\noindent\hrulefill

\noindent\begin{minipage}[b]{\linewidth}
    \lstset{
        language=ttt,
        style=watcode,
        basicstyle=\footnotesize\ttfamily,
        columns=fullflexible,
        breaklines=true}
        \begin{lstlisting}[]{Name}
 RHS (module
    (type (;0;) (func (param i32 f32) (result i64)))
!+_______(func (;0;) (type 0) (param i32 f32) (result i64)!
!+__________i64.const 0) !
        \end{lstlisting}
   \end{minipage}
   
}

\tool never adds a call instruction to this function.
So in practice, the new function is never executed.
Therefore, executing both, the original binary and the mutated one, with the same input, leads to the same final state.
This strategy follows the work of Cohen, advocating the insertion of harmless `garbage' code into a program \cite{cohen1993operating}. 
These transformations increase the static complexity of the generated variant.

\textbf{Remove dead code:} \tool can randomly remove dead code.
In particular, \tool removes: \emph{functions, types, custom sections, imports, tables, memories, globals, data segments, and elements} that can be validated as dead code with guarantees.
For instance, to delete a memory declaration, the binary code must not contain a memory access operation. 
Separated rewriting rules are included within \tool for each of the elements above.
For a more concrete example, the following listing illustrates the case of a function removal.

\lstdefinestyle{watcode}{
  numbers=none,
  stepnumber=1,
  numbersep=10pt,
  tabsize=4,
  showspaces=false,
  breaklines=true, 
  showstringspaces=false,
    moredelim=**[is][{\btHL[fill=weborange!40]}]{`}{`},
    moredelim=**[is][{\btHL[fill=celadon!40]}]{!}{!}
}

{
    \lstset{
        language=ttt,
        style=watcode,
        basicstyle=\footnotesize\ttfamily,
        columns=fullflexible,
        breaklines=true}
         
\vspace{8mm}
        \begin{lstlisting}[]{Name}
LHS (module (type (func)))
        \end{lstlisting}
\noindent\hrulefill
    \lstset{
        language=ttt,
        style=watcode,
        basicstyle=\footnotesize\ttfamily,
        columns=fullflexible,
        breaklines=true}
        \begin{lstlisting}[numbers=none]{Name}
RHS `- (module (import "" "" (func)))`
        \end{lstlisting}

\lstset{
        language=ttt,
        style=watcode,
        basicstyle=\footnotesize\ttfamily,
        columns=fullflexible,
        breaklines=true}
        \begin{lstlisting}[numbers=none]{Name}
Cond The removed function is not called, it is not exported, and it is not in the binary _table.
        \end{lstlisting}
}

In the context of the previous rewriting rule, when removing a function, \tool ensures that the resulting binary remains valid and functionally identical to the original binary: it checks that the deleted function was neither called within the binary code nor exported in the binary external interface. 
As exemplified above, \tool might also eliminate a function import while removing the function. 

Eliminating dead code serves a dual purpose: it minimizes the attack surface available to potential malicious actors \cite{236200} and strengthens the resilience of security protocols. 
For instance, it can obstruct signature-based identification \cite{CABRERAARTEAGA2023103296}.
With Narayan and colleagues having demonstrated the feasibility of Return-Oriented Programming (ROP) attacks \cite{Swivel}, the removal of dead code can stop jumps to harmful behaviors within the binary. 
On the other hand, the act of removing dead code reduces the binary's size, improving its non-functional properties, in particular bandwidth constraints.

\textbf{Edit custom sections:}
Custom sections in WebAssembly are used to store metadata, such as the name of the compiler that produces the binary or the symbol information for debugging.
Thus, this section does not affect the execution of the WebAssembly program.
\tool includes one transformation to edit custom sections. 
This is illustrated in the following rewriting rule.

\lstdefinestyle{watcode}{
  numbers=none,
  stepnumber=1,
  numbersep=10pt,
  tabsize=4,
  showspaces=false,
  breaklines=true, 
  showstringspaces=false,
    moredelim=**[is][{\btHL[fill=weborange!40]}]{`}{`},
    moredelim=**[is][{\btHL[fill=celadon!40]}]{!}{!}
}

{
    \lstset{
        language=ttt,
                        style=watcode,
        basicstyle=\footnotesize\ttfamily,
                        columns=fullflexible,
                        breaklines=true}
        
        \begin{lstlisting}[]{Name}
LHS (module
...
    `-    (@custom "CS42" "zzz..."`
        \end{lstlisting}
\noindent\hrulefill
        
{
    \lstset{
        language=ttt,
                        style=watcode,
        basicstyle=\footnotesize\ttfamily,
                        columns=fullflexible,
                        breaklines=true}
        
        \begin{lstlisting}[]{Name}
RHS (module
...
    !+    (@custom "CS42" "xxx...")!
        \end{lstlisting}
}
}

The \emph{Edit Custom Section} transformation operates by randomly modifying either the content or the name of the custom section. 
As illustrated by Cabrera-Arteaga et al. \cite{CABRERAARTEAGA2023103296}, such a rewriting strategy also acts as a potent deterrent against compiler identification techniques.
Furthermore, it can also be employed innovatively to emulate the characteristics of a different compiler, \emph{masquerading} as another compilation source. 
This strategy ultimately aids in shrinking the identification and fingerprinting surface accessible to potential adversaries, thus enhancing overall system security, or making it a moving target.

\textbf{If swapping:} In WebAssembly, an if-construction consists of a consequence and an alternative. The branching condition is executed right before the \texttt{if} instruction.
If the value at the top of the stack is greater than \texttt{0}, then the consequence code is executed, otherwise the alternative code is run.
The \emph{if swapping} transformation swaps the consequence and alternative codes of an if-construction.


To swap an if-construction in WebAssembly, \tool inserts a negation of the value at the top of the stack right before the \texttt{if} instruction.
In the following rewriting rule, we show how \tool performs this rewriting.

\lstdefinestyle{watcode}{
  numbers=none,
  stepnumber=1,
  numbersep=10pt,
  tabsize=4,
  showspaces=false,
  breaklines=true, 
  showstringspaces=false,
    moredelim=**[is][{\btHL[fill=weborange!40]}]{`}{`},
    moredelim=**[is][{\btHL[fill=celadon!40]}]{!}{!},
    moredelim=**[is][{\btHL[fill=frenchplum!40]}]{bb}{bb},
    moredelim=**[is][{\btHL[fill=eminence!40]}]{bc}{bc}
}

{
    \lstset{
        language=ttt,
                        style=watcode,
        basicstyle=\footnotesize\ttfamily,
                        columns=fullflexible,
                        breaklines=true}
        
        \begin{lstlisting}[]{Name}
LHS (module
    (func ...) (
bb condition C bb
        (if bb A bb  else bb  B bb end)
    )
)
        \end{lstlisting}

\noindent\hrulefill

{
    \lstset{
        language=ttt,
                        style=watcode,
        basicstyle=\footnotesize\ttfamily,
                        columns=fullflexible,
                        breaklines=true}
        
        \begin{lstlisting}[]{Name}
RHS (module
    (func ...) (
bb condition C bb
!i32.eqz!
        (if bb B bb else bb A bb end)
    )
)
        \end{lstlisting}

}
}

The consequence and alternative codes are annotated with the letters \texttt{A} and \texttt{B}, respectively.
The condition of the if-construction is denoted as \texttt{C}.
The negation of the condition is achieved by adding the \texttt{i32.eqz} instruction in the \texttt{RHS} part of the rewriting rule.
The \texttt{i32.eqz} instruction compares the top value of the stack with zero, pushing the value \texttt{1} if the comparison is true.
Some if-constructions may not have either a consequence or an alternative code.
In such cases, \tool replaces the missing code block with a single \texttt{nop} instruction.

\textbf{Loop Unrolling:} 
Loop unrolling is a technique employed to enhance the performance of programs by reducing loop control overhead \cite{dongarra1979unrolling}. 
\tool incorporates a loop unrolling transformation and uses the Abstract Syntax Tree (AST) of the original WebAssembly binary to identify loop constructions. 

When \tool selects a loop for unrolling, its instructions are divided by first-order breaks, which are jumps to the loop's start. This separation ensures that branching instructions controlling the loop body do not require label index adjustments during unrolling. The same holds for instructions continuing to the next loop iteration.
As the loop unrolling process unfolds, a new WebAssembly block is created to encompass both the duplicated loop body and the original loop. 
Within this newly established block, the previously separated groups of instructions are copied. 
These replicated groups of instructions mirror the original ones, except for branching instructions jumping outside the loop body, which need their jumping indices increased by one. This modification is required due to the introduction of a new \texttt{block ... end} scope around the loop body, which affects the scope levels of the branching instructions.

In the following text, we illustrate the rewriting rule for a function that contains a loop. 

\lstdefinestyle{watcode}{
  numbers=none,
  stepnumber=1,
  numbersep=10pt,
  tabsize=4,
  showspaces=false,
  breaklines=true, 
  showstringspaces=false,
    moredelim=**[is][{\btHL[fill=weborange!40]}]{`}{`},
    moredelim=**[is][{\btHL[fill=celadon!40]}]{!}{!},
    moredelim=**[is][{\btHL[fill=frenchplum!40]}]{bb}{bb},
    moredelim=**[is][{\btHL[fill=eminence!40]}]{bc}{bc}
}

{
    \lstset{
        language=ttt,
                        style=watcode,
        basicstyle=\footnotesize\ttfamily,
                        columns=fullflexible,
                        breaklines=true}
        
        \begin{lstlisting}[escapechar=?]{Name}
LHS (module
    (func ...) (
        (loop ? ?   bb A bb  br_if 0 ? ? bc B bc end) ? ?
? ?    )
)
        \end{lstlisting}
 \noindent\hrulefill
       
{
    \lstset{
        language=ttt,
                        style=watcode,
        basicstyle=\footnotesize\ttfamily,
                        columns=fullflexible,
                        breaklines=true}
        
        \begin{lstlisting}[escapechar=?]{Name}
RHS (module
    (func ...) (
        (block
            (block bb A' bb ? ? br_if 0 ? ? bc B' bc ? ? br 1  ? ? end) ? ?
            (loop ? ? bb A' bb ? ? br_if 0 ? ? bc B' bc ? ? end) 
        end) ?\tikzmarkPROBE{8}{bb1}{5}{2}?
    )
)
        \end{lstlisting}
}
}

The loop in the \texttt{LHS} part features a single first-order break, indicating that its execution will cause the program to continue iterating through the loop. 
The loop body concludes right before the \texttt{end} instruction, which highlights the point at which the original loop breaks and resumes program execution.
Upon selecting the loop for unrolling, its instructions are divided into two groups, labeled \texttt{A} and \texttt{B}. 
As illustrated in the \texttt{RHS} part, the unrolling process entails creating two new WebAssembly blocks. 
The outer block encompasses both the original loop structure and the duplicated loop body, while the inner blocks, denoted as \texttt{A'} and \texttt{B'}, represent modifications of the jump instructions in groups \texttt{A} and \texttt{B}, respectively.
Notice that, any jump instructions within \texttt{A'} and \texttt{B'} that originally leaped outside the loop must have their jump indices incremented by one. 
This adjustment accounts for the new block scope introduced around the loop body during the unrolling process. 
Furthermore, an unconditional branch is placed at the end of the unrolled loop iteration's body. 
This ensures that if the loop body does not continue, the tool breaks out of the scope instead of proceeding to the non-unrolled loop.

Loop unrolling enhances resistance to static analysis while maintaining the original performance \cite{10.1145/3453483.3454035}. 
In particular, Crane et al. \cite{10.1145/2810103.2813682} have validated the effectiveness of adding and modifying jump instructions against Function-Reuse attacks.
Our rewriting rule has the same advantages, it unrolls loops while 1) incorporating new jumps and 2) editing existing jumps, as it can be observed with the addition of the \texttt{br_if}, \texttt{end}, and \texttt{br} instructions.

\textbf{Peephole:} 
This meta-rule is about rewriting peephole instruction sequences within function bodies, signifying the most granular level of rewriting. 
We implement 125 rewriting rules for this group in \tool. 
We include rewriting rules that affect the memory of the binary.
For example, we include rewriting rules that create random assignments to newly created global variables.
For these rules, we incorporate several conditions, denoted by \texttt{Cond}, to ensure successful replacement. 
These conditions can be used interchangeably and combined to constrain transformations.

For instance, \tool is designed to guarantee that instructions marked for replacement are deterministic. 
We specifically exclude instructions that could potentially cause undefined behavior, such as function calls, from being mutated. 
For this rewriting type, \tool only alters stack and memory operations, leaving the control frame labels unaffected.
\revision{In the following, we show 14 peephole rewriting rules.}

\begin{revision1}
    
    We include well-known identity reduction rewriting rules, we exemplify them in the following.
Observe that the primitive type of the rewriting rules can be changed to the other primitive \Wasm data types. 

\begin{minipage}{0.95\linewidth}
\begin{minipage}{0.49\linewidth}
    
    \lstset{
    language=ttt,
    style=watcode,
    basicstyle=\footnotesize\ttfamily,
    columns=fullflexible,
    breaklines=true}
    \begin{lstlisting}[]
LHS i32.or ?x i32.const.-1
            \end{lstlisting}\vspace{-0.5cm}
    \noindent\hrulefill
        \lstset{
            language=ttt,
            style=watcode,
            basicstyle=\footnotesize\ttfamily,
            columns=fullflexible,
            breaklines=true}
            \vspace{-0.2cm}
            \begin{lstlisting}[numbers=none]{Name}
RHS i32.const.-1
    \end{lstlisting}
\end{minipage}
\begin{minipage}{0.49\linewidth}
    \lstset{
    language=ttt,
    style=watcode,
    basicstyle=\footnotesize\ttfamily,
    columns=fullflexible,
    breaklines=true}
    \begin{lstlisting}[]
LHS f32.mul ?x f32.const.1
            \end{lstlisting}\vspace{-0.5cm}
    \noindent\hrulefill
        \lstset{
            language=ttt,
            style=watcode,
            basicstyle=\footnotesize\ttfamily,
            columns=fullflexible,
            breaklines=true}
            \vspace{-0.2cm}
            \begin{lstlisting}[numbers=none]{Name}
RHS ?x
    \end{lstlisting}
\end{minipage}    
\end{minipage}

\begin{minipage}{0.95\linewidth}
\begin{minipage}{0.49\linewidth}
    
    \lstset{
    language=ttt,
    style=watcode,
    basicstyle=\footnotesize\ttfamily,
    columns=fullflexible,
    breaklines=true}
    \begin{lstlisting}[]
LHS i32.eq ?x i32.const.0
            \end{lstlisting}\vspace{-0.5cm}
    \noindent\hrulefill
        \lstset{
            language=ttt,
            style=watcode,
            basicstyle=\footnotesize\ttfamily,
            columns=fullflexible,
            breaklines=true}
            \vspace{-0.2cm}
            \begin{lstlisting}[numbers=none]{Name}
RHS i32.eqz ?x
    \end{lstlisting}
\end{minipage}
\begin{minipage}{0.49\linewidth}
    \lstset{
    language=ttt,
    style=watcode,
    basicstyle=\footnotesize\ttfamily,
    columns=fullflexible,
    breaklines=true}
    \begin{lstlisting}[]
LHS i64.shl ?x i64.const.0
            \end{lstlisting}\vspace{-0.5cm}
    \noindent\hrulefill
        \lstset{
            language=ttt,
            style=watcode,
            basicstyle=\footnotesize\ttfamily,
            columns=fullflexible,
            breaklines=true}
            \vspace{-0.2cm}
            \begin{lstlisting}[numbers=none]{Name}
RHS ?x
    \end{lstlisting}
\end{minipage}    
\end{minipage}

We have also incorporated commutative rewriting rules. 
The following three rules provide examples. 
As discussed earlier, observe that the subexpressions' type can be changed to any of the original four primitive \Wasm data types.

\begin{minipage}{0.95\linewidth}
\begin{minipage}{0.49\linewidth}
    
    \lstset{
    language=ttt,
    style=watcode,
    basicstyle=\footnotesize\ttfamily,
    columns=fullflexible,
    breaklines=true}
    \begin{lstlisting}[]
LHS select ?y ?y ?x
            \end{lstlisting}\vspace{-0.5cm}
    \noindent\hrulefill
        \lstset{
            language=ttt,
            style=watcode,
            basicstyle=\footnotesize\ttfamily,
            columns=fullflexible,
            breaklines=true}
            \vspace{-0.2cm}
            \begin{lstlisting}[numbers=none]{Name}
RHS ?y
    \end{lstlisting}
\end{minipage}
\begin{minipage}{0.49\linewidth}
    \lstset{
    language=ttt,
    style=watcode,
    basicstyle=\footnotesize\ttfamily,
    columns=fullflexible,
    breaklines=true}
    \begin{lstlisting}[]
LHS i32.add ?x ?y
            \end{lstlisting}\vspace{-0.5cm}
    \noindent\hrulefill
        \lstset{
            language=ttt,
            style=watcode,
            basicstyle=\footnotesize\ttfamily,
            columns=fullflexible,
            breaklines=true}
            \vspace{-0.2cm}
            \begin{lstlisting}[numbers=none]{Name}
RHS i32.add ?y ?x
    \end{lstlisting}
\end{minipage}    
\end{minipage}

\begin{minipage}{0.95\linewidth}
\begin{minipage}{0.49\linewidth}
    
    \lstset{
    language=ttt,
    style=watcode,
    basicstyle=\footnotesize\ttfamily,
    columns=fullflexible,
    breaklines=true}
    \begin{lstlisting}[]
LHS i32.mul ?x ?y
            \end{lstlisting}\vspace{-0.5cm}
    \noindent\hrulefill
        \lstset{
            language=ttt,
            style=watcode,
            basicstyle=\footnotesize\ttfamily,
            columns=fullflexible,
            breaklines=true}
            \vspace{-0.2cm}
            \begin{lstlisting}[numbers=none]{Name}
RHS i32.mul ?y ?x
    \end{lstlisting}
\end{minipage}
\begin{minipage}{0.49\linewidth}
\end{minipage}    
\end{minipage}

The peephole meta-rule also includes associative rewriting rules.
In the following, we exemplify two of them.

\begin{minipage}{0.95\linewidth}
\begin{minipage}{0.49\linewidth}
    
    \lstset{
    language=ttt,
    style=watcode,
    basicstyle=\footnotesize\ttfamily,
    columns=fullflexible,
    breaklines=true}
    \begin{lstlisting}[]
LHS i32.mul ?x (i32.mul ?y ?z)
            \end{lstlisting}\vspace{-0.5cm}
    \noindent\hrulefill
        \lstset{
            language=ttt,
            style=watcode,
            basicstyle=\footnotesize\ttfamily,
            columns=fullflexible,
            breaklines=true}
            \vspace{-0.2cm}
            \begin{lstlisting}[numbers=none]{Name}
RHS i32.mul (i32.mul ?x ?y) ?z
    \end{lstlisting}
\end{minipage}
\begin{minipage}{0.49\linewidth}
    \lstset{
    language=ttt,
    style=watcode,
    basicstyle=\footnotesize\ttfamily,
    columns=fullflexible,
    breaklines=true}
    \begin{lstlisting}[]
LHS i32.add ?x (i32.add ?y ?z)
            \end{lstlisting}\vspace{-0.5cm}
    \noindent\hrulefill
        \lstset{
            language=ttt,
            style=watcode,
            basicstyle=\footnotesize\ttfamily,
            columns=fullflexible,
            breaklines=true}
            \vspace{-0.2cm}
            \begin{lstlisting}[numbers=none]{Name}
RHS i32.add (i32.add ?x ?y) ?z
    \end{lstlisting}
\end{minipage}    
\end{minipage}

We have also incorporated several strength reduction transformations.
We exemplify them in the following three rewriting rules.
Note that the extent of these rewriting rules can be adjusted as desired.
For instance, although we only include reductions up to multiplication by 8 (in the rightmost rewriting rule), they can be extended to any power of 2.

\begin{minipage}{0.95\linewidth}
\begin{minipage}{0.49\linewidth}
    
    \lstset{
    language=ttt,
    style=watcode,
    basicstyle=\footnotesize\ttfamily,
    columns=fullflexible,
    breaklines=true}
    \begin{lstlisting}[]
LHS i32.shl ?x i32.const.1
            \end{lstlisting}\vspace{-0.5cm}
    \noindent\hrulefill
        \lstset{
            language=ttt,
            style=watcode,
            basicstyle=\footnotesize\ttfamily,
            columns=fullflexible,
            breaklines=true}
            \vspace{-0.2cm}
            \begin{lstlisting}[numbers=none]{Name}
RHS i32.mul ?x i32.const.2
    \end{lstlisting}
\end{minipage}
\begin{minipage}{0.49\linewidth}
    \lstset{
    language=ttt,
    style=watcode,
    basicstyle=\footnotesize\ttfamily,
    columns=fullflexible,
    breaklines=true}
    \begin{lstlisting}[]
LHS i32.shl ?x i32.const.3
            \end{lstlisting}\vspace{-0.5cm}
    \noindent\hrulefill
        \lstset{
            language=ttt,
            style=watcode,
            basicstyle=\footnotesize\ttfamily,
            columns=fullflexible,
            breaklines=true}
            \vspace{-0.2cm}
            \begin{lstlisting}[numbers=none]{Name}
RHS i32.mul ?x i32.const.8
    \end{lstlisting}
\end{minipage}    
\end{minipage}

\begin{minipage}{0.95\linewidth}
\begin{minipage}{0.49\linewidth}
    
    \lstset{
    language=ttt,
    style=watcode,
    basicstyle=\footnotesize\ttfamily,
    columns=fullflexible,
    breaklines=true}
    \begin{lstlisting}[]
LHS i32.add ?x ?x
            \end{lstlisting}\vspace{-0.5cm}
    \noindent\hrulefill
        \lstset{
            language=ttt,
            style=watcode,
            basicstyle=\footnotesize\ttfamily,
            columns=fullflexible,
            breaklines=true}
            \vspace{-0.2cm}
            \begin{lstlisting}[numbers=none]{Name}
RHS i32.mul ?x i32.const.2
    \end{lstlisting}
\end{minipage}
\begin{minipage}{0.49\linewidth}
\end{minipage}    
\end{minipage}

As previously mentioned, our work is built on well-established diversification strategies. 
In the subsequent rewriting rules, we exemplify the porting of two well-known strategies. 
The leftmost part illustrates the injection of \texttt{nop} instructions \cite{10.1145/2086696.2086702}. 
The rightmost part extend the peephole meta-rule with the "un-folding" of constants rewriting rule. 
In the latter case, statically defined constants are substituted by the sum of two numbers. 
This sum computes the original constant at runtime.

\begin{minipage}{0.95\linewidth}
\begin{minipage}{0.49\linewidth}
    
    \lstset{
    language=ttt,
    style=watcode,
    basicstyle=\footnotesize\ttfamily,
    columns=fullflexible,
    breaklines=true}
    \begin{lstlisting}[]
LHS ?x
            \end{lstlisting}\vspace{-0.5cm}
    \noindent\hrulefill
        \lstset{
            language=ttt,
            style=watcode,
            basicstyle=\footnotesize\ttfamily,
            columns=fullflexible,
            breaklines=true}
            \vspace{-0.2cm}
            \begin{lstlisting}[numbers=none]{Name}
RHS (nop ?x)
    \end{lstlisting}
\end{minipage}
\begin{minipage}{0.49\linewidth}
    \lstset{
    language=ttt,
    style=watcode,
    basicstyle=\footnotesize\ttfamily,
    columns=fullflexible,
    breaklines=true}
    \begin{lstlisting}[]
LHS ?x
            \end{lstlisting}\vspace{-0.5cm}
    \noindent\hrulefill
        \lstset{
            language=ttt,
            style=watcode,
            basicstyle=\footnotesize\ttfamily,
            columns=fullflexible,
            breaklines=true}
            \vspace{-0.2cm}
            \begin{lstlisting}[numbers=none]{Name}
RHS i32.add (i32.const z i32.const y) 
    \end{lstlisting}\vspace{-0.5cm}
    \noindent\hrulefill
\lstset{
        language=ttt,
        style=watcode,
        basicstyle=\footnotesize\ttfamily,
        columns=fullflexible,
        breaklines=true}
        \begin{lstlisting}[numbers=none]{Name}
Cond z = i32 random & y = x - z 
        \end{lstlisting}
\end{minipage}    
\end{minipage}

The implemented rewriting rules can be employed commutatively. 
For example, the left-hand side (LHS) can function interchangeably as the right-hand side (RHS), and the reverse is also valid. 
Consequently, observe the doubling of practical rewriting occurrences. 
The same logic applies to numeric types, i.e., some rewriting rules could be duplicated and then changed to another \Wasm data type.
    
\end{revision1}


\subsection{E-graphs for WebAssembly}
\label{alg}

We build \tool on top of e-graphs \cite{10.1145/3571207}.
An e-graph is a graph data structure used for representing rewriting rules and their chaining. 
In an e-graph, there are two types of nodes: e-nodes and e-classes. 
An e-node represents either an operator or an operand involved in the rewriting rule, while an e-class denotes the equivalence classes among e-nodes by grouping them, i.e., an e-class is a virtual node compound of a collection of e-nodes. 
Thus, e-classes contain at least one e-node.
Edges within the graph establish operator-operand equivalence relations between e-nodes and e-classes.

In \tool, the e-graph is automatically built from a WebAssembly program by analyzing its expressions and operations through its data flow graph.
Then, each unique expression, operator, and operand are transformed into e-nodes.
Based on the input rewriting rules, the equivalent expressions are detected, grouping equivalent e-nodes into e-classes.
During the detection of equivalent expressions, new operators could be added to the graph as e-nodes.
Finally, e-nodes within an e-class are connected with edges to represent their equivalence relationships.

\begin{figure}
    \centering
    \includegraphics[width=1.0\linewidth]{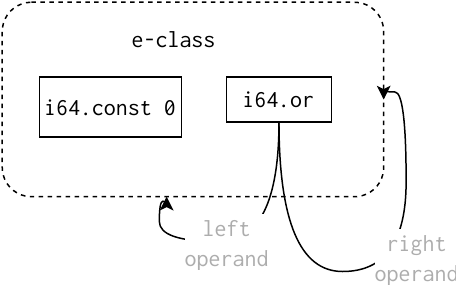}
    \caption{e-graph for idempotent bitwise-or rewriting rule. Solid lines represent operand-operator relations, and dashed lines represent equivalent class inclusion. }
  \label{e-graph}
\end{figure}

For example, let us consider one program with a single instruction that returns an integer constant, \texttt{i64.const 0}. Let us also assume a single rewriting rule, \texttt{(x,\ x\ i64.or\ x, x instanceof i64)}. 
In this example, the program's control flow graph contains just one node, representing the unique instruction.
The rewriting rule represents the equivalence for performing an \texttt{or} operation with two equal operands.
\autoref{e-graph} displays the final e-graph data structure constructed out of this single program and rewriting rule. 
We start by adding the unique program instruction \texttt{i64.const 0} as an e-node (depicted by the leftmost solid rectangle node in the figure). 
Next, we generate e-nodes from the rewriting rule (the rightmost solid rectangle) by introducing a new e-node, \texttt{i64.or}, and creating edges to the \texttt{x} e-node.
Following this, we establish equivalence. 
The rewriting rule combines the two e-nodes into a single e-class (indicated by the dashed rectangle node in the figure). 
As a result, we update the edges to point to the \texttt{x} symbol e-class.


Willsey et al. illustrate that the extraction of code fragments from e-graphs can achieve a high level of flexibility, especially when the extraction process is recursively defined through a cost function applied to e-nodes and their operands. 
This approach guarantees the semantic equivalence of the extracted code \cite{10.1145/3434304}. 
For example, to obtain the smallest code from an e-graph, one could initiate the extraction process at an e-node and then choose the AST with the smallest size from among the operands of its associated e-class \cite{10.1145/3385412.3386012}.
When the cost function is omitted from the extraction methodology, the following property emerges:
\emph{Any path traversed through the e-graph will result in a semantically equivalent code variant}. 
This concept is illustrated in \autoref{e-graph}, where it is possible to construct an infinite sequence of "or" operations.
In the current study, we leverage this inherent flexibility to generate mutated variants of an original program. 
The e-graph offers the option for random traversal, allowing for the random selection of an e-node within each e-class visited, thereby yielding an equivalent expression.


 

\algnewcommand\algorithmicforeach{\textbf{for each}}
\algdef{S}[FOR]{ForEach}[1]{\algorithmicforeach\ #1\ \algorithmicdo}
\begin{algorithm}
    \footnotesize
	\begin{algorithmic}[1]
        \Procedure{traverse}{$e-graph$, $eclass$, $depth$}
        \If{depth = 0}
          \State  \Return \textbf{smallest\_tree\_from}(e-graph,\ eclass)
        \Else
            \State $nodes \gets e-graph[eclass]$
            \State $node \gets random\_choice(nodes)$
            \State $expr \gets (node, operands=[])$
            \ForEach {$child \in node.children $}
                \State $subexpr \gets \textbf{TRAVERSE}(e-graph,\ child,\ depth - 1)$
                \State $expr.operands \gets expr.operands \cup\ \{subexpr\}$
            \EndFor
            \State \Return $expr$
        \EndIf
        \EndProcedure
	\end{algorithmic} 
	\caption{e-graph traversal algorithm.} 
	\label{peephole:mutator}
\end{algorithm}

We propose and implement an algorithm to randomly traverse an e-graph and generate semantically equivalent program variants, see Algorithm \autoref{peephole:mutator}.
It receives an e-graph, an e-class node (initially the root's e-class), and the maximum depth of expression to extract. The depth parameter ensures that the algorithm is not stuck in an infinite recursion. We select a random e-node from the e-class (lines 5 and 6), and the process recursively continues with the children of the selected e-node (line 8) with a decreasing depth. As soon as the depth becomes zero, the algorithm returns the smallest expression out of the current e-class (line 3). The subexpressions are composed together (line 10) for each child, and then the entire expression is returned (line 11). 
To the best of our knowledge, \tool, is the first practical implementation of random e-graph traversal for \Wasm.

\begin{figure}
    \centering
    \includegraphics[width=0.9\linewidth]{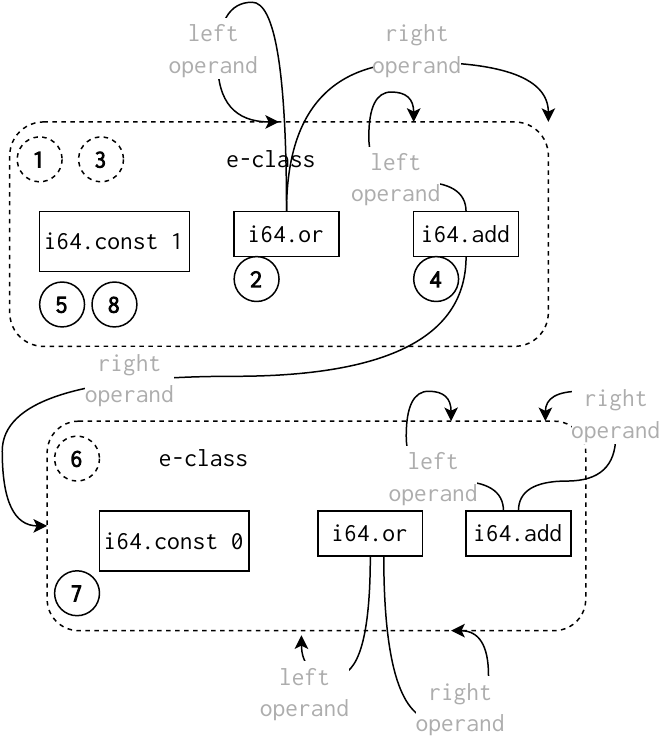}
    \caption{e-graph built for rewriting the first instruction of \autoref{example:peeporig}. }
  \label{e-graph3}
\end{figure}

Let us demonstrate how the proposed traversal algorithm can generate program variants with an example. 
We will illustrate Algorithm \ref{peephole:mutator} using a maximum depth of 1. 
\autoref{example:peeporig} presents a hypothetical original WebAssembly binary to mutate. 
In this example, the developer has established two rewriting rules: \texttt{(x, x i32.or x, x instanceof i32)} and \texttt{(x, x i32.add 0, x instanceof i32)}. The first rewriting rule represents the equivalence of performing an \texttt{or} operation with two equal operands, while the second rule signifies the equivalence of adding 0 to any numeric value.
By employing the code and the rewriting rules, we can construct the e-graph depicted in \autoref{e-graph3}. The figure demonstrates the operator-operand relationship using arrows between the corresponding nodes.

\lstdefinestyle{watcode}{
  numbers=none,
  stepnumber=1,
  numbersep=10pt,
  tabsize=4,
  showspaces=false,
  breaklines=true, 
  showstringspaces=false,
    moredelim=**[is][{\btHL[fill=weborange!40]}]{`}{`},
    moredelim=**[is][{\btHL[fill=celadon!40]}]{!}{!}
}

   \begin{minipage}[b]{\linewidth}
    \lstset{
        language=WAT,
                        style=watcode,
        basicstyle=\footnotesize\ttfamily,
                        columns=fullflexible,
                        breaklines=true}
        
        \begin{lstlisting}[label=example:peeporig,caption={Wasm function.},frame=b, captionpos=b]{Name}
(module
    (type (;0;) (func (param i32 f32) (result i64)))
    (func (;0;) (type 0) (param i32 f32) (result i64)
        i64.const 1)
)
        \end{lstlisting}
\end{minipage}

\begin{minipage}[b]{\linewidth}
    \lstset{
        language=WAT,
                        style=watcode,
        basicstyle=\footnotesize\ttfamily,
                        columns=fullflexible,
                        breaklines=true}
        
        \begin{lstlisting}[label=example:peepapplied,caption={Random peephole mutation using egraph traversal for \autoref{example:peeporig} over e-graph \autoref{e-graph3}. The textual format is folded for better understanding.},frame=b, captionpos=b]{Name}
(module
    (type (;0;) (func (param i32 f32) (result i64)))
    (func (;0;) (type 0) (param i32 f32) (result i64)
        !(i64.or (!
            !(i64.add (!
                !i64.const 0!
                !i64.const 1!
            !))!
            !i64.const 1!
        !))!
    )
        \end{lstlisting}
\end{minipage}

In \autoref{e-graph3}, we annotate the various steps of Algorithm \ref{peephole:mutator} 
for the scenario described above. Algorithm \ref{peephole:mutator} begins at the e-class containing the single instruction \texttt{i64.const 1} from \autoref{example:peeporig}. 
It then selects an equivalent node in the e-class \step{2}, in this case, the \texttt{i64.or} node, resulting in:
{\texttt{expr = i64.or l r}}.
The traversal proceeds with the left operand of the selected node \step{3}, choosing the \texttt{i64.add} node within the e-class: 
{\texttt{expr = i64.or (i64.add l r)} \texttt{r}}.
The left operand of the \texttt{i64.add} node is the original node \step{5}: 
{\texttt{expr = i64.or (i64.add i64.const 1 r)} \texttt{r}}.
The right operand of the \texttt{i64.add} node belongs to another e-class, where the node \texttt{i64.const 0} is selected \step{6}\step{7}:
{\texttt{expr = i64.or (i64.add i64.const 1 i64.const 0)} \texttt{r}}.
In the final step \step{8}, the right operand of the \texttt{i64.or} is selected, corresponding to the initial instruction e-node, returning:
{\texttt{expr = i64.or (i64.add i64.const 1 i64.const 0)\ i64.const 1}}
The traversal result applied to the original WebAssembly code can observed in \autoref{example:peepapplied}.

\begin{revision1}
    
\subsection{\tool in the loop}
In practice, we use \tool as a crucial component within a wider process. 
This process includes an additional validation and verification of the generated variants. 
The general process starts with a \wasm binary as the input and iterates over the variants created by \tool to offer guarantees. 
These guarantees are ensured through five specific components: 1) e-graphs traversals by \tool, 2) a static validation of the variants, 3) a dynamic verification of the variant's state after execution by the host engine, 4) the assessment of different machine code according to the JIT engine, and 5) the detection of unique execution traces during the run. 
Notice that the composition of these components minimizes the number of variants that \tool can generate without functional equivalence, due to the potential incorrect implementation of the tool. 
For example, if the \emph{Remove Dead Code} meta-rule implementation wrongly removes an imported function, which is later called, the variant fails the static validation. 
In Algorithm \autoref{toolinpractice}, we illustrate how \tool plays a role in a wider process.

The algorithm starts by executing the original \wasm program and recording its original state, as indicated in line 5.
This initial state serves as a reference for validating and evaluating subsequent variants.
The state includes 1) the linear memory after the program initialization (e.g., by invoking the \texttt{_start} function of the initial \wasm binary if it exists), 2) the value and type of the global variables, and 3) the standard output of the program.

We start \tool in line 8 of the algorithm, initiating the loop for creating a new variant.
If a variant is generated out of \tool in line 8, static validation is invoked in line 9.
This static validation is performed by the wasmtime parser \footnote{https://crates.io/crates/wasmparser}.
The static validation step ensures the stack's soundness.
For instance, the static validator verifies that a function call does not refer to a non-existent function index.
It also ensures the correctness of the execution stack, i.e., a non-returning block should leave the stack in the same state as before execution.

Upon successful static validation of a unique variant, line 11 triggers a JIT compilation within the \wasm engine.
This step compiles the variant into machine code.
The algorithm then checks whether this machine code differs from the original, thus confirming diversity at the machine code level.
If this condition is met, we execute the variant to collect its state.

Then, the variant's state is compared to the original state for dynamic validation.
This process is demonstrated in line 14.
We perform a fine-grained comparison of the variant's globals, linear memory, and output according to a validation workload.
For example, the variant and the original are considered identical for the linear memory if all bytes are equal and in the same order.
If there is any difference in these state components, the variant is discarded because it is not equivalent to the oriugina

The loop ends when a unique variant with new traces distinct from the original is found, as validated in line 17.
The algorithm then returns the generated variant, ensuring that both diversified machine code and traces differ from the original.
If this condition is not met, the transformation is stacked in line 19, and the loop restarts.

\algdef{S}[FOR]{ForEach}[1]{\algorithmicforeach\ #1\ \algorithmicdo}
\begin{algorithm*}{
	\begin{algorithmic}[1]
        \Procedure{diversify}{$originalWasm$, $engine$}
        \\
         Input:
        \Comment{A \wasm binary to diversify and a \wasm engine.}
        \\
         Output:
        \Comment{A statically unique and behaviourally different \wasm variant.}
        \\
        \State $originalState \gets \textbf{engine.execute}(originalWasm)$
        
        \State $wasm \gets originalWasm$
        \While{true}
             \State $variantWasm \gets {\color{darkgreen}\textbf{WASM-MUTATE}(wasm)}$
             \If{{\color{darkgreen}staticValidate($variantWasm$)}}
                 \If{ $variantWasm$ is unique}                    
                    \State $variantJIT \gets \textbf{engine.compile}(variantWasm)$
                    \If{$variantJIT$ is unique}
                        \State $state \gets \textbf{engine.execute}(variantJIT)$
                        \If{{\color{darkgreen} $state.memory == originalState.memory$  $\&\&$ \\ 
                                            $state.globals == originalState.globals$ $\&\&$ \\ 
                                            $state.output == originalState.output$}}
        
                            \If{$state.trace \neq originalState.trace$ }
                               \State \Return $variantWasm$
                            \EndIf
                            \State $wasm \gets variantWasm$ // we stack the transformation
                        \EndIf
                    \EndIf
                 \EndIf
             \EndIf
        \EndWhile
        \EndProcedure
	\end{algorithmic} 
	\caption{\tool in the loop.} 
	\label{toolinpractice}
}
\end{algorithm*}
\end{revision1}

\subsection{Implementation}

\tool is implemented in Rust, comprising approximately, 10000 lines of Rust code. 
We leverage the capabilities of the wasm-tools project of the bytecode alliance for parsing and transforming WebAssembly binary code. 
Specifically, we use the wasmparser, \url{https://github.com/bytecodealliance/wasm-tools/tree/main/crates/wasmparser} and wasm-encoder, \url{https://github.com/bytecodealliance/wasm-tools/tree/main/crates/wasm-encoder} modules for parsing and encoding WebAssembly binaries, respectively.
\revision{
The wasmparser crate provides quick, efficient decoding and parsing of WebAssembly binary files.
Its primary advantage is a minimal memory footprint, achieved without creating AST or IR of WebAssembly data.
Conversely, the wasm-encoder crate is a reliable library for encoding Wasm binaries.
Additionally, the wasm-encoder crate offers static validation of the constructed binary, ensuring the integrity of a newly encoded Wasm binary.
}
The implementation of \tool is publicly available for future research and can be found at \repourl.


\section {Evaluation}
\label{eval}

In this section, we outline our methodology for evaluating \tool.
Initially, we introduce our research questions and the corpora of programs that we use for the assessment of \tool.
Next, we elaborate on the methodology for each research question.
For the sake of reproducibility, our data and experimenting pipeline are publicly available at \dataurl.
Our experiments are conducted in Standard F4s-v2(Skylake) Azure machines with 4 virtual CPUs and 8GiB memory per instance.


\newcommand\rqstatic{To what extent are the program variants generated by \tool statically different from the original programs?\xspace}

\newcommand\rqdynamic{How fast can \tool generate program variants that exhibit different execution traces?\xspace}

\newcommand\rqdefensive{To what extent does \tool prevent side-channel attacks on \Wasm programs?\xspace}

\newcommand\rqperformance{To what extent does \tool affect the performance of real-world \Wasm program variants?\xspace}

\newcommand\rqtesting{To what extent can \tool be used to perform differential testing of \Wasm tools?\xspace}

\newcommand{\nProgramsRosetta}{303\xspace}

\newcommand{\DTWStatic}{\ensuremath{\mathit{dt\_static}\xspace}}
\newcommand{\DTWDynamic}{\ensuremath{\mathit{dt\_dy}\xspace}}

\begin{enumerate}[label=RQ\arabic*:, ref=RQ\arabic*]
     \item \label{rq:static} \textbf{\rqstatic}
        We check whether the \wasm binary variants rapidly produced by \tool are different from the original \wasm binary. Then, we assess whether the x86 machine code produced by the wasmtime engine is also different.
    
    \item \label{rq:dynamic}\textbf{\rqdynamic}
    To assess the versatility of \tool, we also examine the presence of different behaviors in the generated variants. 
    Specifically, we measure the speed at which \tool generates variants with distinct machine code instruction traces and memory access patterns.
    

     \begin{revision1}
        \item \label{rq:performance}\textbf{\rqperformance} This research question evaluates the performance impact of \tool in creating variants from real-world programs. We compare the machine code size and the execution time of the original programs and their variants.
     \end{revision1}
     
    \item \label{rq:defensive}\textbf{\rqdefensive} Software Diversification is an option to prevent security issues. In this research question we assess the impact of \tool in preventing one class of attacks: cache timing attacks (Spectre).

\end{enumerate}

\subsection{Corpora}
\label{sec:corpus}

\begin{table}
\renewcommand\arraystretch{1.1}
\begin{adjustbox}{width=\linewidth,totalheight=\textheight, keepaspectratio}
{
    \begin{tabular}{p{1.5cm} | l | p{1cm} | l| p{2cm}  }
        \hline
        Source & Programs & RQ & \ Mean \# Ins. & Note  \\
        \hline \hline
        
        CROW \cite{arteaga2020crow} & 303 prog. & \ref{rq:static}, \ref{rq:dynamic} &  8451 & Rosetta  \\
        \hline

        wasmbench \cite{hilbig2021empirical} & 134 prog. & \ref{rq:performance} & 12665 & Real world \\
        
        \hline

        Swivel \cite{Swivel} & 2 prog. & \ref{rq:defensive} & 297;743 & Spectre BTB  \\
        \hline
        Safeside \cite{Swivel, safeside}  & 2 prog. & \ref{rq:defensive} & 378894;379058 & Spectre RSB \& Spectre PHT  \\

    \end{tabular}
}
\end{adjustbox}
    
    \caption{ \revision{
        Dataset of 441 programs that we use to evaluate \tool. Each row in the table corresponds to programs, with the columns providing: where the program is sourced from, the number of programs, research question addressed,  the mean number of instructions found in the original \wasm program and, a short note about the programs.}}
    \label{tab:corpus}
\end{table}


We use a collection of programs comprised of four curated corpora to address our research questions. 
Our corpora contains a total \revision{ of 441 programs (303 + 134 + 2 + 2).}
The metadata of these programs is summarized in \autoref{tab:corpus}.
Each row in the table corresponds to the programs in use for each research question. 
The columns provide information on the following: the corpus source, the number of programs, the research question addressed, and the mean number of instructions for the programs. 
Additionally, a short note highlights the main property of the programs in the last column. 
For instance, the notes in the last two rows indicate the attacks to which the programs are susceptible.

We answer \ref{rq:static} and \ref{rq:dynamic} with a corpus of programs from Cabrera \etal \cite{arteaga2020crow}, which is shown in the first row of \autoref{tab:corpus}.
The corpus contains \nProgramsRosetta programs.
The programs in the corpus perform a range of tasks, from simple ones, such as sorting, to complex algorithms like a compiler lexer. 
The number of total instructions ranges from 170 to 36023 with a mean of 8451 instructions.
All programs in the corpus: 
1) do not require input from users, \ie do not functions like \texttt{scanf}, 2) terminate, 3) are deterministic, \ie given the same input, provide the same output and 4) compile to \wasm using \texttt{wasi-clang} to compile them.
The size of the binaries ranges from 465 to 92114 bytes. 

\begin{revision1}
To address \ref{rq:performance}, we evaluate the performance impact of the \tool variants on 134 real-world programs. 
We gather these programs from the wasmbench dataset \cite{hilbig2021empirical}. 
The wasmbench dataset consists of 8461 \Wasm binaries, which were gathered from the internet in 2021. 
For our experiment, we need binaries that can execute without user interaction and that do not rely on external resources such as complementary JavaScript code. 
We filter wasmbench to retrieve all the binaries with \texttt{_start} functions that can be executed directly with wasmtime+WASI in less than 60 seconds. 
This provides us with  134 real-world \Wasm binaries, which can be used to assess the performance implications of diversification. 
As visible in the second row of \autoref{tab:corpus}, the programs are large, with 12665 instructions as the mean value.
The size of the binaries ranges from 113 to 4304430 bytes. 
\end{revision1}

We address \ref{rq:defensive} with four \Wasm programs and three Spectre attack scenarios sourcing from the Swivel and Safeside projects \cite{Swivel,safeside}. 
The specifics of these programs are revealed in the final two rows of \autoref{tab:corpus}.
The first two programs, containing 297 and 743 instructions respectively, are intentionally designed for the Spectre branch target attack (BTB).
These programs have 954 and 1910 bytes in size. 
The last two programs, presented in the final row, are derived from the Safeside project \cite{safeside}. 
Unlike the first pair, these programs are substantially larger, containing more than 300000 instructions each and having sizes exceeding 1500000 bytes. 
They are used to execute the Spectre Return Stack (RSB) and Spectre Pattern History (PHT) attacks \cite{Spectre}. 
The significant difference in terms of the number of instructions and sizes of the first two programs and the last pair is due to the contrasting compilation processes used for these \Wasm binaries. 
A detailed description of the three attack scenarios is provided in \autoref{protocol:rq3}.

\subsection{Protocol for RQ1}
\label{protocol:rq1}

\begin{figure}
    \centering
    \includegraphics[width=0.8\linewidth]{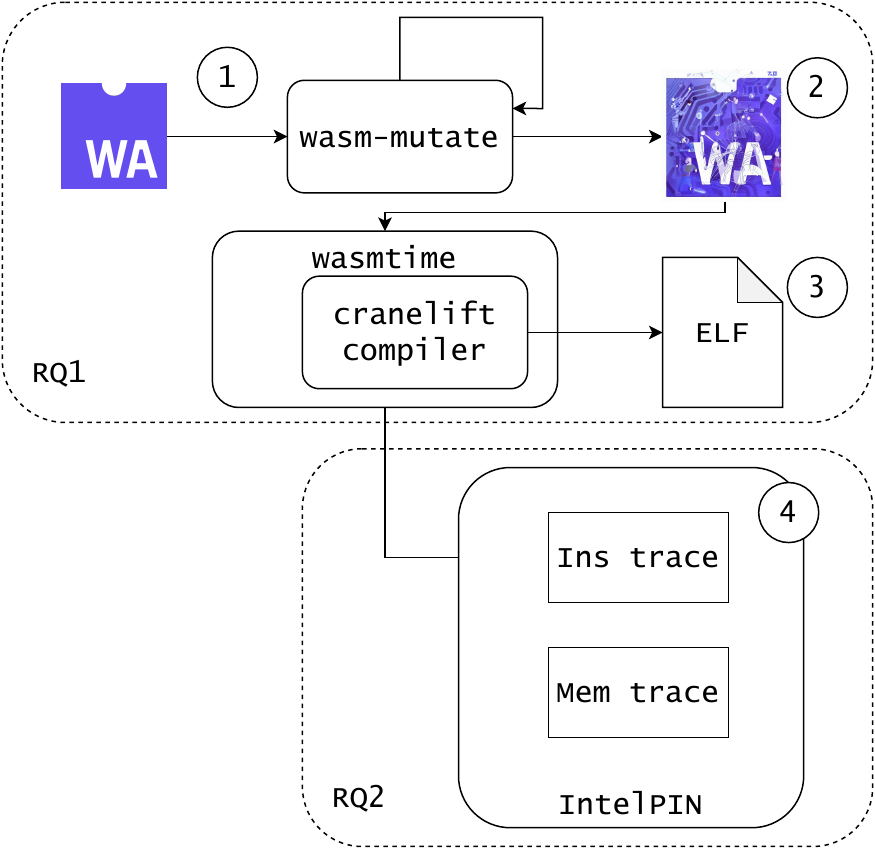}
    \caption{Protocol to answer \ref{rq:static} and \ref{rq:dynamic}}
  \label{protocol}
\end{figure}

With \ref{rq:static},
we assess the ability of \tool to generate \Wasm variants that are statically different from the original program, including after their compilation to x86 machine code.
In \autoref{protocol} we show the steps we follow to answer \ref{rq:static}.
We run \tool on the corpus of \nProgramsRosetta{} original C programs described in the first row of \autoref{tab:corpus} (step \step{1} in figure). 
To generate the variants:
1) we start with one original program and pass it to \tool to generate a variant;  
2) the variants and the original program form a population of variants; 
3) we randomly select a program from this population and pass it to \tool to generate another variant, which we add to the population; 
4) we then restart the process in the previous step to stack more mutations 
This procedure is carried out for 1 hour.
The outcome (step \step{2} in the figure) is a population containing programs with many stacked transformations, all starting from an original \wasm program.
We then count the number of unique variants in the population.
We compute the sha256 hash of each variant bytestream in order to calculate the population size.
We define the population size metric as:

\begin{metric}{Population\_size(P):}\label{metric:pop}
Given an original \wasm program P, a generated corpus of \wasm programs $V=\{v_1, v_2, ..., v_N\}$ where $v_i$ is a variant of P, the population size is defined as:
$$
    | set(\{ sha256(v_1), ... sha256(v_N) \})|\text{ }\forall v_i \in V 
$$
\end{metric}

Since \wasm binaries may be further transformed into machine code before they execute, we also check that these additional transformations preserve the difference introduced by \tool in the \wasm binary. 
We use the wasmtime JIT compiler, Cranelift, with all available optimizations, to generate the x86 binaries for each \wasm program and its variants  (step \step{3} in figure). 
Then, we calculate the number of unique variants of machine code representation for wasmtime.
Counting the number of unique machine codes, we compute the diversification preservation ratio: \\

\begin{metric}{Ratio of preserved variants:}\label{metric:preservation}
    Given an original \wasm program P and its population size as defined in \autoref{metric:pop} and the JIT compiler C, we defined the ratio of preserved variants as:
    $$
        \frac{ | set(\{ sha256(C(v_1)), ... sha256(C(v_N)) \})|}{ \text{Population\_size (P)}} \text{ }\forall v_i \in V 
    $$

\end{metric}

If  $sha256(P_1)$ $\neq$ $sha256(P_2)$ and $sha256(C(P_1))$) $\neq$ $sha256(C(P_2))$, this means that both programs are still different after being compiled to machine code, and this means that the Cranelift compiler has not removed the transformations made by \tool.  

Note that the protocol described earlier can be mapped to Algorithm \autoref{toolinpractice}. 
For instance, to measure population size for each tested program, one could measure how often the execution of Algorithm \autoref{toolinpractice} reaches line 11. 
Similarly, to assess the level of preservation, one could track the frequency with which the algorithm arrives at line 13.

\subsection{Protocol for RQ2}
\newcommand{\samples}{100\xspace}

For \ref{rq:dynamic}, we evaluate how fast \tool can generate variants that offer distinct traces compared with the original program.
We start by collecting the traces of the original program when executed in wasmtime.
While continuously generating variants with randomly stacked transformations, we collect the execution traces of the variants as well.
We record the time passed until we generate a variant that offers different execution traces, according to two types of traces: machine code instructions and memory accesses.
This process can be seen in the enclosed square of \autoref{protocol}, annotated with \ref{rq:dynamic}.

We gather the instructions and memory traces utilizing IntelPIN \cite{luk2005pin, 10.1145/3478520} (step \step{4} in the figure).
To only collect the traces of the \Wasm execution with a wasmtime engine, we pause and resume the collection as the execution leaves and re-enters the \Wasm code, respectively.
We implement this filtering with the built-in hooks of wasmtime.
In addition, we disable ASLR on the machine where the variants are executed.
This latter action ensures that the placement of the instructions in memory is deterministic.
Examples of the traces we collect can be seen in \autoref{example:trace1} and \autoref{example:trace2} for memory and instruction traces, respectively.

\lstdefinestyle{watcode}{
  numbers=none,
  stepnumber=1,
  numbersep=10pt,
  tabsize=4,
  showspaces=false,
  breaklines=true, 
  showstringspaces=false,
    moredelim=**[is][{\btHL[fill=weborange!40]}]{`}{`},
    moredelim=**[is][{\btHL[fill=celadon!40]}]{!}{!}
}

   \begin{minipage}[b]{0.9\linewidth}
    \lstset{
        language=TRACE,
                        style=watcode,
        basicstyle=\footnotesize\ttfamily,
                        columns=fullflexible,
                        breaklines=true}
        
        \begin{lstlisting}[label=example:trace1,caption={Memory trace  with two events out of IntelPIN for the execution of a \wasm program with wasmtime. Trace events record: the type of the operation, read or write, the memory address, the number of bytes affected and the value read or written.},frame=b, captionpos=b]{Name}
[Writ]                                                           0x555555ed1570 size=4 value=0x10dd0
[Read]                                                           0x555555ed1570 size=4 value=0x10dd0                      \end{lstlisting}
\end{minipage}

\begin{minipage}[b]{0.9\linewidth}
    \lstset{
        language=TRACE,
                        style=watcode,
        basicstyle=\footnotesize\ttfamily,
                        columns=fullflexible,
                        breaklines=true}
        
        \begin{lstlisting}[label=example:trace2,caption={Instructions trace with two events out of IntelPIN for the execution of a \wasm program with wasmtime. Each event records the corresponding machine code that executes.},frame=b, captionpos=b]{Name}
[I]          mov rdx, qword ptr [r14+0x100]           
[I]          mov dword ptr [rdx+0xe64], ecx          
    \end{lstlisting}
\end{minipage}

In the text below, we outline the metric used to assess how fast \tool can generate variants that provide different execution traces.

\begin{metric}{Time until different trace:}\label{metric:mem:sha}
Given an original \wasm program P, and its execution trace $T_1$, the time until a different trace is defined as the time between the diversification process starts and when the variant $V$ is generated with execution trace $T_2$ with $T_1 \neq T_2$.
\\

Notice that the previously defined metric is instantiated twice, for instructions and memory type of events.

\end{metric}

Referring to Algorithm \ref{toolinpractice}, we quantify the elapsed time between line 6 and line 18 to obtain the time it takes for \tool to generate a unique \Wasm variant producing different execution traces.

\begin{revision1}

\subsection{Protocol for \ref{rq:performance}}
We evaluate the performance implications of using \tool to generate variants. 
To do this, we use programs from previous work \cite{hilbig2021empirical}. 
We collect the subset of 134 real-world programs that can be executed with no input and no user interaction.
For each program in the corpus, we generate a maximum of 50 unique variants, with every variant comprising 1000 stacked transformations. 
The performance of these variants is determined by measuring both the binary size of the compiled program and the execution time for each program and its corresponding variants. 
For binary size, we collect the binary size after native code compilation by wasmtime.
In total, we measure these parameters across a set of 6834 (134 + $134\times50$) programs. 
To compare all programs and their variants, we apply the following relative metrics:

\begin{metric}{Relative Machine Code size impact:}\label{metric:mcize} Given an original \Wasm program P and a variant V, given MP and MV the machine code for P and V respectively, the relative machine code size is defined as:

$$
    \frac{|MV|}{|MP|}
$$

Here, the \textit{| |} function returns the byte size of an input binary.     
\end{metric}

Finally, we execute each program and variant and measure their execution time. 
We then compare the execution time of variants for its original program execution time.

\begin{metric}{Relative execution time:}\label{metric:ex}
Given a program P and a variant V, the relative execution time is defined as:

$$
    \frac{init(V)}{init(P)}
$$

\texttt{init} measures the time of executing the \texttt{_start} function of an already JITed \Wasm binary.
\end{metric}

We collect \autoref{metric:ex} after running each program and variant 100 times.
We discard the first 20 measurements to remove noise and warm up the JIT engine.
Thus, for each program and variant, we collect 80 execution times.
  
\end{revision1}

\subsection{Protocol for \ref{rq:defensive}}
\label{protocol:rq3}

\newcommand{\poct}{\emph{Cache timing POC}\xspace}
\newcommand{\pocd}{\emph{Differential computing POC}\xspace}
\newcommand{\pocp}{\emph{Port contention POC}\xspace}

To answer \ref{rq:defensive}, we apply \tool to the same security \wasm programs used by Narayan et al. to evaluate Swivel's ability to protect \wasm programs against side-channel attacks \cite{Swivel}. 
The four cache timing side-channel attacks are presented in detail in \autoref{sec:corpus}. 
The specific binary and its corresponding attack can be appreciated in \autoref{tab:corpus}.
We evaluate to what extent \tool can prevent such attacks.
In the following text, we describe the attacks we replicate and evaluate in order to answer \ref{rq:defensive}.


Narayan and colleagues successfully bypass the control flow integrity safeguards of \Wasm, using speculative code execution as detailed in \cite{Spectre}. 
Thus, we use the same three Spectre attacks from Swivel:
1) The Spectre Branch Target Buffer (btb) attack exploits the branch target buffer by predicting the target of an indirect jump, thereby rerouting speculative control flow to an arbitrary target.
2) The Spectre Pattern History Table (pht) takes advantage of the pattern history table to anticipate the direction of a conditional branch during the ongoing evaluation of a condition. 
3) The Spectre Return Stack Buffer (ret2spec) attack exploits the return stack buffer that stores the locations of recently executed call instructions to predict the target of \texttt{ret} instructions. 
Each attack methodology relies on the extraction of memory bytes from another hosted \wasm binary that executes in parallel.

For each of the four \wasm binaries introduced in \autoref{sec:corpus}, we generated a maximum of 1000 random stacked transformations utilizing 100 distinct seeds. 
This resulted in a total of 100,000 variants for each original \wasm binary.
We then assess the success rate of attacks across these variants by measuring the bandwidth of the exfiltrated data, that is: the rate of correctly leaked bytes per unit of time. 
We then count the correctly exfiltrated bytes and divided them by the variant program's execution time. 

Notice that, the bandwidth metric captures not only whether the attacks are successful or not, but also the degree to which the data exfiltration is hindered.
For instance, a variant that continues to exfiltrate secret data but does so over an impractical duration would be deemed as having been hardened. 
For this, we state the bandwidth metric in the following definition :

\begin{metric}{Attack bandwidth:}\label{metric:ber}
Given data $D=\{b_0, b1, ..., b_C\}$ being exfiltrated in time $T$ and $K = {k_1, k_2, ..., k_N}$ the collection of correct data bytes, the bandwidth metric is defined as:
$$
    \frac{|b_i\text{ such that } b_i \in K|}{T}
$$
\end{metric}

\section{Experimental Results}
\label{results}

\subsection{\rqstatic}
\label{rq:static:results}

\newcommand{\preserved}{62\%\xspace}

To address \ref{rq:static}, we use \tool to process the original 303 programs from \cite{arteaga2020crow}. 
\tool is set to generate variants with a timeout of one hour for each program. 
Following this, we assess the sizes of their variant populations as well as their corresponding preservation ratio (Refer to \autoref{metric:pop} and \autoref{metric:preservation} for more details).

\begin{figure}
    \centering
    \includegraphics[width=\linewidth]{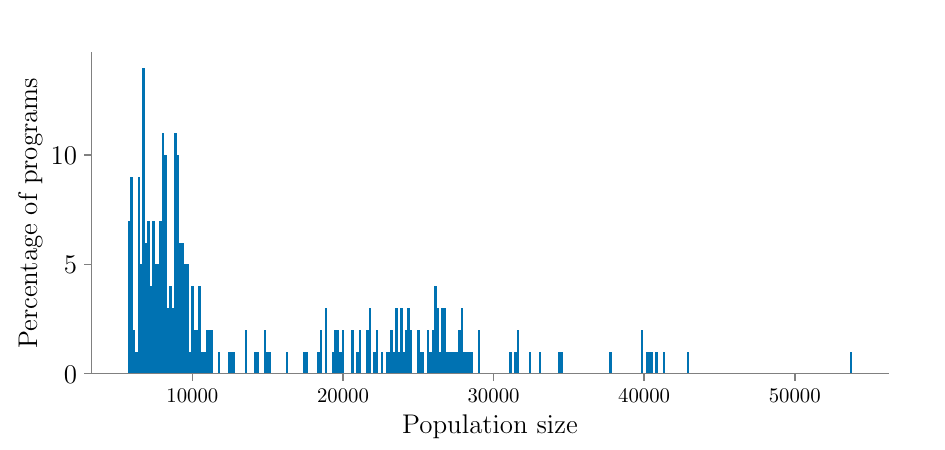}
    \caption{RQ1: Number of unique \wasm programs generated by \tool in 1 hour for each program of the corpus.}
  \label{rq1:plot:population}
\end{figure}

In \autoref{rq1:plot:population}, we show the distribution of the population size generated out of \tool.
\tool successfully diversifies all 303 original programs, yielding a diversification rate of 100\%. 
Within an hour, \tool demonstrates its impressive efficiency and effectiveness by producing a median of 9500 unique variants for the 303 original programs.
The largest population size observed is 53816, while the smallest is 5716.
There are several factors contributing to large population sizes. 

\tool can diversify functions within WASI-libc. 
Despite the relatively low function count in the original source code,  \tool creates thousands of distinct variants in the function of the incorporated libraries. 
This feature improves over methods that can only diversify the original source code processed through the LLVM compilation pipeline \cite{arteaga2020crow}. 


We have observed a significant variation in the population size out of \tool between different programs, ranging by several thousand variants (from a maximum of 53816 variants to a minimum of 5716 variants).
This disparity is attributed to:
the non-deterministic nature of \tool and 2) the characteristics of the program. 
\tool mutates a randomly selected portion of a program. 
If the selected instruction is determined to be non-deterministic, despite the transformation being semantically equivalent, \tool discards the variant and moves on to another random transformation.
For instance, if the instruction targeted for mutation is a function call, \tool proceeds to the next one.
This process, in conjunction with the unique characteristics of each program, results in a varying population size. 
For example, an input binary with a high number of function calls would lead to a greater number of trials and errors, slowing down the generation of variants, thereby resulting in a smaller overall population size for 1 hour of \tool execution.



\begin{figure}
    \centering
    \includegraphics[width=\linewidth]{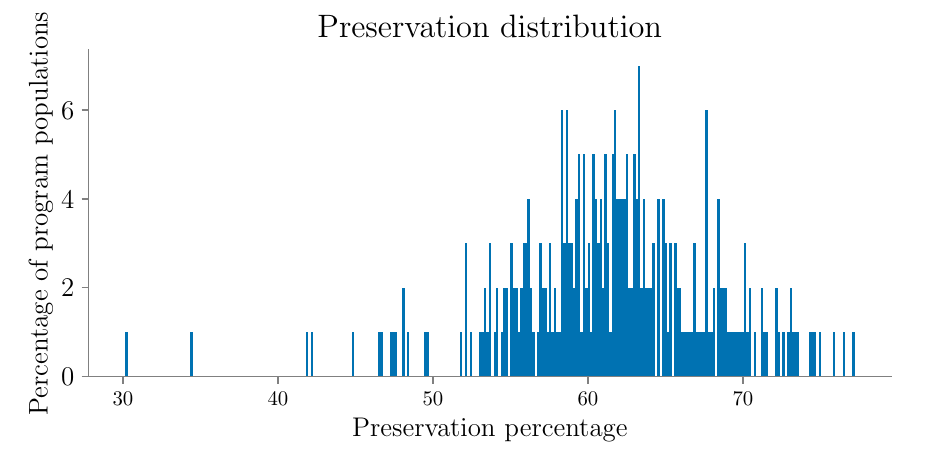}
    \caption{RQ1: Distribution of the ratio of wasmtime preserved variants.}
  \label{rq1:plot:preservation}
\end{figure}

As stated in \autoref{protocol:rq1}, we also assess static diversification with \autoref{metric:preservation} by calculating the preservation ratio of variant populations. 
\autoref{rq1:plot:preservation} presents the distribution of preservation ratios for the Cranelift compiler of wasmtime. 
We have observed a median preservation ratio of \preserved. 
On the one hand, we have observed that there is no correlation between population size and preservation ratio. 
In other words, having a larger population size does not necessarily lead to a higher preservation ratio.
On the other hand, the phenomena of non-preserved variants can be explained as follows. 
Factors such as custom sections are often disregarded by compilers. 
Similarly, bloated code plays a role in this context. 
For instance, \tool generates certain variants with unused types or functions, which are then detected and eliminated by Cranelift.
Yet, note that even when working with the smallest population size and the lowest preservation percentage, the number of unique machine codes can still encompass thousands of variants.




\begin{tcolorbox}[boxrule=1pt,arc=.3em,boxsep=-1.3mm]
  \textbf{Answer to \ref{rq:static}}: \tool generates \wasm variants for all the 303 input programs. 
  Within a one-hour diversification budget, \tool synthesizes more than 9000 unique variants per program on average. 
  \preserved of the variants remain different after machine-code compilation.
  \tool is good at producing many \Wasm program variants.
\end{tcolorbox}

\subsection{\rqdynamic}

\begin{figure}
    \centering
    \includegraphics[width=\linewidth]{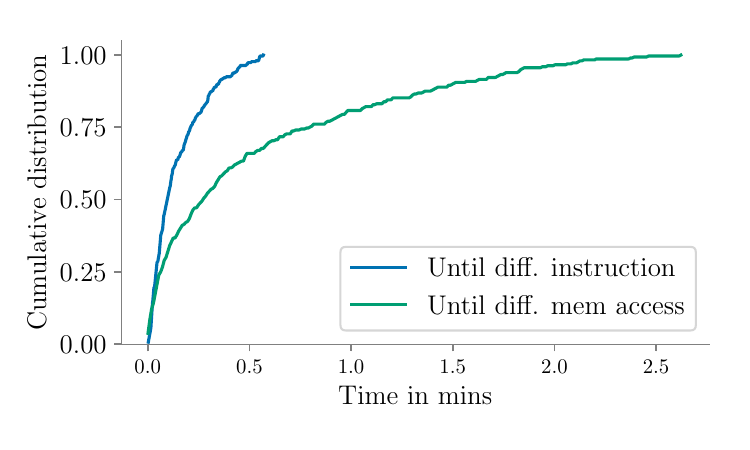}
    \caption{RQ2: Cumulative distribution for the time until different trace. In blue for different machine code instructions, in green for different memory traces. The X-axis marks time in minutes, and the Y-axis shows the ratio of programs from \nProgramsRosetta for which \tool created a variant within that time.}
  \label{rq2:plot:mem}
\end{figure}

To answer question \ref{rq:dynamic},   we measure how long it takes to generate one variant that exhibits execution traces that are different from the original.
In \autoref{rq2:plot:mem}, we display a cumulative distribution plot showing the time required for \tool to generate variants with different traces, in blue for machine code instructions and green for memory traces.
The X-axis marks time in minutes, and the Y-axis shows the ratio of programs from \nProgramsRosetta for which \tool created a variant within that time.
For all original programs, \tool succeeds in  generating one variant with different traces compared to the original program, either in machine code instructions or memory access, i.e., both cumulative distributions reach 100\%
The shortest time to generate a variant with different machine code instruction traces is 0.12 seconds, and for different memory traces, it is 0.06 seconds. 
In the slowest scenarios, \tool takes under 1 minute for different machine code instruction traces and less than 3 minutes for different memory traces.
Overall, \tool takes a median of 5.4 seconds and 12.6 seconds in generating variants with different machine code instructions and different memory instructions respectively.

The use of an e-graph random traversal is the key factor for such a fast generation process.
Once \tool locates a modifiable instruction within the binary and constructs its corresponding e-graph, traversal is virtually instantaneous. 
However, the time efficiency of variant generation is not consistent across all programs, as illustrated in \autoref{rq2:plot:mem}. 
This variation primarily stems from the varying complexities of the programs under analysis, as previously mentioned in \autoref{rq:static:results}.
Interestingly, \tool may attempt to build e-graphs from instructions that, while not inherently leading to undefined behavior, are part of a data flow graph that could. 
For example, the data flow graph might be dependent on a function call. 
Although transforming undefined behavioral instructions is deactivated by default in \tool to maintain functional equivalence with the original code, the process of attempting to construct such e-graphs can extend the duration of the diversification pass.
As a result, \tool may require multiple attempts to successfully create and traverse an e-graph, impacting the rate at which it generates behaviorally distinct variants. 
This phenomenon is particularly noticeable in original programs that have a high frequency of function calls.

On average, \tool takes three times longer to synthesize unique memory traces than it does to generate different instruction traces (as can be observed in how the green plot of the figure is skewed to the right). 
The main reason for this difference is the limited set of rewriting rules that specifically focus on memory operations. 
\tool includes more rules for manipulating code, which increases the odds of generating a variant with diverse machine code instructions.
Additionally, the variant creation process halts and restarts with alternative rewriting rules if \tool detects that the selected code for transformation could result in unpredictable behavior. 

We have identified four primary factors explaining why execution traces differ overall.
First, alterations to the binary layout inherently impact both machine code instruction traces and memory accesses within the program's stack. 
In particular, \tool creates variants that change the return addresses of functions, leading to divergent execution traces, including those related to memory access. 
Second, our rewriting rules incorporate artificial global values into \wasm binaries. 
Since these global variables are inherently manipulated via the stack, their access inevitably generates divergent memory traces.
Third, \tool injects 'phantom' instructions which do not aim to modify the outcome of a transformed function during execution. 
These intermediate calculations trigger the spill/reload component of the runtime, varying spill and reload operations. 
In the context of limited physical resources, these operations temporarily store values in memory for later retrieval and use, thus creating unique memory traces.
Finally, certain rewriting rules implemented by \tool replicate fragments of code, e.g., performing commutative operations. 
These code segments may contain memory accesses, and while neither the memory addresses nor their values change, the frequency of these operations does.
Overall, these findings influence the diversity of execution traces among the generated variants. 

\begin{tcolorbox}[boxrule=1pt,arc=.3em,boxsep=-1.3mm]
  \textbf{Answer to \ref{rq:dynamic}}: \tool generates variants with distinct machine code instructions and memory traces for all tested programs. 
  The quickest time for generating a variant with a unique machine code trace is 0.12 seconds, and for divergent memory traces, the fastest generation only lasts 0.06 seconds. 
  On average, the median time required to produce a variant with distinct traces stands at 5.4 seconds for different machine code traces and 16.2 seconds for different memory traces. 
  These metrics indicate that \tool is suitable for fast-moving target defense strategies, capable of generating a new variant in well under a minute \cite{MEWE}. To the best of our knowledge, \tool is the fastest diversification engine for \wasm.
\end{tcolorbox}

\begin{revision1}
\subsection{\rqperformance}

To answer \ref{rq:performance}, we generate variants for 134 real-world programs selected from the wasmbench dataset \cite{hilbig2021empirical}.
We produce 50 variants for each original program and execute them along with the original program, for a total of $(50 + 1)\times134$ variants.
We collect the machine code size after the \Wasm program is JITed by wasmtime and the execution time of each original program/variant.

\begin{figure}
    \centering
    \includegraphics[width=\linewidth]{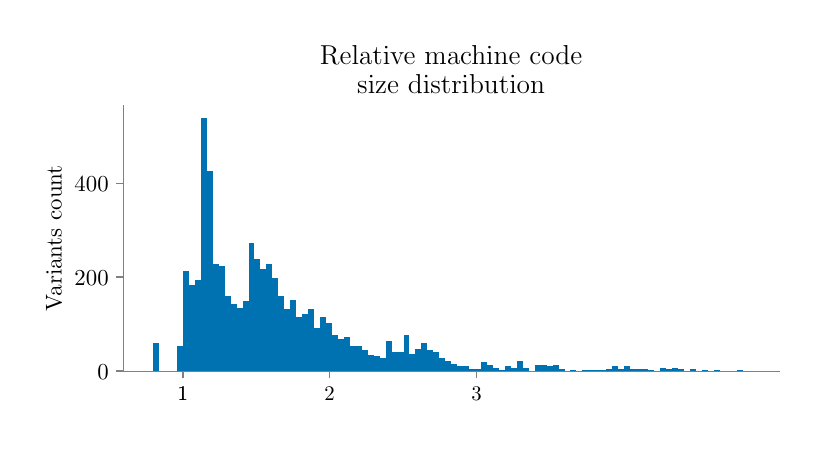}
    \caption{Distribution of the relative machine code size.
The Y-axis represents the count of variants, while the X-axis represents the relative size of the variants, in the number of machine code bytes. \tool generates \Wasm variants that exhibit a broad range of binary sizes.}
  \label{impact:mcsize}
\end{figure} 

\begin{figure}
    \centering
    \includegraphics[width=\linewidth]{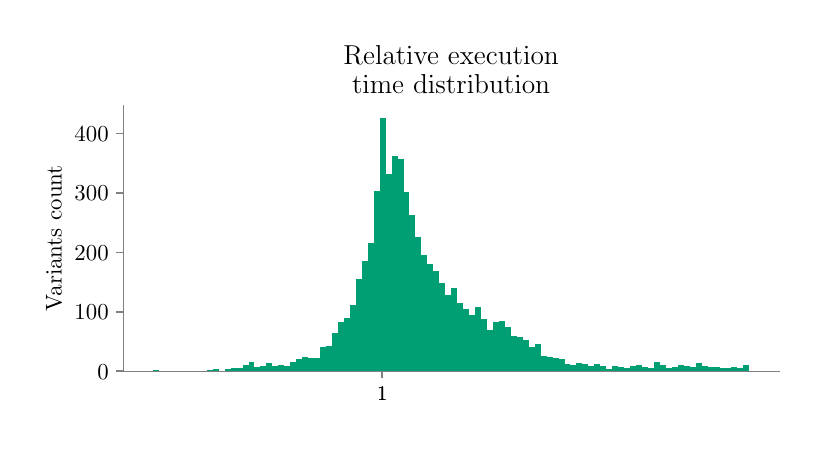}
    \caption{Distribution of relative execution times for variants produced by \tool.
The Y-axis represents the count of variants, while the X-axis represents the relative execution times of the variants. Relative execution times display a normal distribution, 29\% of the variants are faster while 70\% operate slower. }
  \label{impact:time}
\end{figure}

In \autoref{impact:mcsize}, we plot the relative machine code size distribution.
The Y-axis represents the count of variants, while the X-axis represents the relative value of the relative machine code size.
The median relative machine code size is 1.6.
The size of the generated machine has a maximum increase of 4.8 times the original size.
Overall, the wasmtime compiler yields a diverse range of machine code sizes.
The larger size of the produced machine code could potentially affect the distribution of programs across the network.
Therefore, practitioners may opt to distribute only those variants that only have a small increase in binary size when bandwidth presents a significant concern.
Also, we highlight that 2\% of the generated variants result in a smaller machine code size, showing that our diversification engine generates variants that are optimized.

In \autoref{impact:time} we show the distribution of relative execution times for variants produced by \tool.
The Y-axis represents the count of variants, while the X-axis represents the relative value of the relative execution time size.
The median relative execution time is 1.1.
Overall, we observe a Gaussian-like shape.
We note that 29\% of the variants perform faster than the original, in contrast to 70\% that are slightly slower.
In the worst-case scenario, \tool produces variants that perform twice as slowly as the original.
On the other hand, in the best-case scenario, \tool produces variants that are five times faster than the original.
We have identified two primary reasons for the minimal impact of \tool on variant execution times.
First, \tool generates variants by injecting phantom code which, in practice, is not executed.
This impacts the machine code size, but not the execution time.
Second, \tool produces variants by optimizing the original programs, in particular, due to peep-hole diversification techniques.
This allows for the generation of faster variants through the stacking of optimizations during the e-graph traversal by \tool.

\vspace{-0.1cm}

\begin{tcolorbox}[boxrule=1pt,arc=.3em,boxsep=-1.3mm]
  \textbf{Answer to \ref{rq:performance}}: 
    \revision{
    \tool generates \Wasm variants that tend to be larger than the original program. 
    Execution times for \tool variants display a normal distribution close to the original performance.
    In general, the impact of \tool remains within the same order of magnitude with a median value of 1.1 in the original execution time. 
    The diversification workflow of \tool can be expanded by practitioners to filter out extreme variants, such as the ones that are too big or too slow. 
    }
\end{tcolorbox}

\end{revision1}

\subsection{\rqdefensive}

\begin{figure*}
    \centering
    \includegraphics[width=\linewidth]{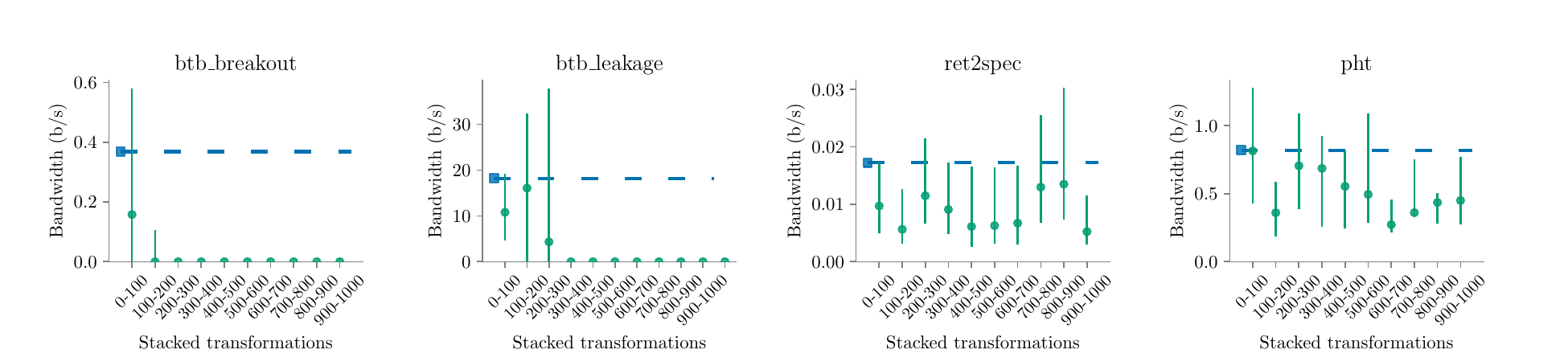}
    \caption{Visual representation of \tool's impact on Swivel's original programs. The Y-axis denotes exfiltration bandwidth, with the original binary's bandwidth under attack highlighted by a blue marker and dashed line. Variants are clustered in groups of 100 stacked transformations, denoted by green dots (median bandwidth) and lines (interquartile bandwidth range). Overall, for all 100000 variants generated out of each original program, 70\% have less data leakage bandwidth.}
  \label{attacks:impact}
\end{figure*}


To answer \ref{rq:defensive}, we execute \tool on four distinct binaries \wasm susceptible to Spectre-related attacks. Each of the four programs is transformed with one of 100 different seeds and up to 1000 stacked transformations. 
We assess the resulting impact of the attacks as outlined in \ref{protocol:rq3}. 
The analysis encompasses a total of 4$\times$100$\times$1000 binaries, which also includes the original four.

\autoref{attacks:impact} offers a graphical representation of \tool's influence on the Swivel original programs and their attacks. 
Each plot corresponds to one original \wasm binary and the attack it undergoes: btb\_breakout, btb\_leakage, ret2spec, and pht.
The Y-axis represents the exfiltration bandwidth (see \autoref{metric:ber}). 
The bandwidth of the original binary under attack is marked as a blue dashed horizontal line.
In each plot, the variants are grouped in clusters of 100 stacked transformations. 
These are indicated by green dots and lines. 
The dot signifies the median bandwidth for the cluster, while the line represents the interquartile range of the group's bandwidth.

For btb\_breakout and btb\_leakage, \tool demonstrates effectiveness, generating variants that leak less information than the original in 78\% and 70\% of the cases, respectively.
For these particular binaries, a significant reduction in exfiltration bandwidth to zero is noted after 200 stacked transformations.
This means that with a minimum of 200 stacked transformations, \tool can create variants that are completely resistant to the original attack.
For the ret2spec and pht scenarios, the produced variants consistently exhibit lower bandwidth than the original in 76\% and 71\% of instances, respectively.
As depicted in the plots, the exfiltration bandwidth diminishes following the application of at least  100 stacked transformations.

This success is explained by the fact that \tool synthesizes variants that effectively alter memory access patterns. 
Specifically, it does so by amplifying spill/reload operations, injecting artificial global variables, and changing the frequency of pre-existing memory accesses. 
These transformations influence the \wasm program's memory, disrupting cache predictors. 

Many attacks rely on a timer component to measure cache access time for memory, and disrupting this component effectively impairs the attack's effectiveness. 
This strategy of dynamic alteration has also been employed in other scenarios. 
For instance, to counter potential timing attacks, Firefox randomizes its built-in JavaScript timer \cite{10.1007/978-3-319-70972-7_13}. \tool applies the same strategy by interspersing instructions within the timing steps of \wasm variants. 
In \autoref{example:timer} and \autoref{example:timer2}, we demonstrate \tool's impact on time measurements. 
The former illustrates the original time measurement, while the latter presents a variant with \tool-inserted operations amid the timing.

\lstdefinestyle{watcode}{
  numbers=none,
  stepnumber=1,
  numbersep=10pt,
  tabsize=4,
  showspaces=false,
  breaklines=true, 
  showstringspaces=false,
    moredelim=**[is][{\btHL[fill=weborange!40]}]{`}{`},
    moredelim=**[is][{\btHL[fill=celadon!40]}]{!}{!}
}

   \begin{minipage}[b]{0.9\linewidth}
    \lstset{
        language=WAT,
                        style=watcode,
        basicstyle=\footnotesize\ttfamily,
                        columns=fullflexible,
                        breaklines=true}
        
        \begin{lstlisting}[label=example:timer,caption={Wasm timer used in btb\_breakout program.},frame=b, captionpos=b]{Name}
;; Code from original btb_breakout
...
(call $readTimer)
(set_local $end_time)
... access to mem
(i64.sub (get_local $end_time ) (get_local $start_time))
(set_local $duration)
...

        \end{lstlisting}
\end{minipage}

\begin{minipage}[b]{0.9\linewidth}
    \lstset{
        language=WAT,
                        style=watcode,
        basicstyle=\footnotesize\ttfamily,
                        columns=fullflexible,
                        breaklines=true}
        
        \begin{lstlisting}[label=example:timer2,caption={Variant of btb\_breakout with more instructions added in between time measurement.},frame=b, captionpos=b]{Name}
;; Variant code
...
(call $readTimer)
(set_local $end_time)
!<inserted instructions>!
... access to mem
!<inserted instructions>!
(i64.sub (get_local $end_time ) (get_local $start_time))
(set_local $duration)
...
        \end{lstlisting}
\end{minipage}

\tool proves effective against cache access timers because the time measurement of a single or a few instructions is inherently different. 
By introducing more instructions, this randomness is amplified, thereby reducing the timer's accuracy.

Furthermore, CPUs have a maximum capacity for the number of instructions they can cache.
\tool injects instructions in such a way that the vulnerable instruction may exceed this cacheable instruction limit, meaning that caching becomes disabled.
This kind of transformation can be viewed as padding \cite{padding}.
In \autoref{example:padding} and \autoref{example:padding2}, we illustrate the effect of \tool on padding instructions.
\autoref{example:padding} presents the original code used for training the branch predictor, along with the expected speculated code.

\lstdefinestyle{watcode}{
  numbers=none,
  stepnumber=1,
  numbersep=10pt,
  tabsize=4,
  showspaces=false,
  breaklines=true, 
  showstringspaces=false,
    moredelim=**[is][{\btHL[fill=weborange!40]}]{`}{`},
    moredelim=**[is][{\btHL[fill=celadon!40]}]{!}{!}
}

   \begin{minipage}[b]{0.9\linewidth}
    \lstset{
        language=WAT,
                        style=watcode,
        basicstyle=\footnotesize\ttfamily,
                        columns=fullflexible,
                        breaklines=true}
        
        \begin{lstlisting}[label=example:padding,caption={Two jump locations in btb\_breakout. The top one trains the branch predictor, the bottom one is the expected jump that exfiltrates the memory access.},frame=b, captionpos=b]{Name}
;; Code from original btb_breakout
...
;; train the code to jump here (index 1)
(i32.load (i32.const 2000))
(i32.store (i32.const 83)) ;; just prevent optimization
...
;; transiently jump here
(i32.load (i32.const 339968)) ;; S(83) is the secret
(i32.store (i32.const 83)) ;; just prevent optimization
        \end{lstlisting}
\end{minipage}

\begin{minipage}[b]{0.9\linewidth}
    \lstset{
        language=WAT,
                        style=watcode,
        basicstyle=\footnotesize\ttfamily,
                        columns=fullflexible,
                        breaklines=true}
        
        \begin{lstlisting}[label=example:padding2,caption={Variant of btb\_breakout with more instructions added indindinctly between jump places.},frame=b, captionpos=b]{Name}
;; Variant code
...
;; train the code to jump here (index 1)
!<inserted instructions>!
(i32.load (i32.const 2000))
!<inserted instructions>!
(i32.store (i32.const 83)) ;; just prevent optimization
...
;; transiently jump here
!<inserted instructions>!
(i32.load (i32.const 339968)) ;; "S"(83) is the secret
!<inserted instructions>!
(i32.store (i32.const 83)) ;; just prevent optimization
...
        \end{lstlisting}
\end{minipage}

The padding alters the arrangement of the binary code in memory, effectively impeding the attacker's capacity to initiate speculative execution.
Even when an attack is launched and the vulnerable code is "speculated", the memory access is not impacted as planned.

In every program, we note that the exfiltration bandwidth tends to be greater than the original when the variants include a few transformations.
This indicates that, although the transformations generally contribute to the reduction of data leakage, the initial few might not consistently contribute positively towards this objective.
We have identified several fundamental reasons, which we discuss below.

First, as emphasized in prior applications of \tool \cite{CABRERAARTEAGA2023103296}, uncontrolled diversification can be counterproductive if a specific objective, such as a cost function, is not established at the beginning of the diversification process.
Secondly, while some transformations yield distinct \wasm binaries, their compilation produces identical machine code.
Transformations that are not preserved undermine the effectiveness of diversification.
For example, incorporating random \texttt{nop} operations directly into \wasm does not modify the final machine code as the \texttt{nop} operations are often removed by the compiler.
The same phenomenon is observed with transformations to custom sections of \Wasm binaries.
Additionally, it is important to note that transformed code doesn't always execute, i.e., \tool may generate dead code.

Finally, for ret2spec and pht, both programs are hardened with attack bandwidth reduction, but this does not materialize in a short-term timeframe (low count of stacked transformations).
Furthermore,  the exfiltration bandwidth is more dispersed for these two programs.
Our analysis indicates a correlation between bandwidth reduction and the complexity of the binary subject to diversification.
Ret2spec and pht are considerably larger than btb\_breakout and btb\_leakage.
The former comprises more than 300k instructions, while the latter two include fewer than 800 instructions.
Given that \tool applies precise, fine-grained transformations one at a time, the likelihood of impacting critical attack components, such as timing memory accesses, diminishes for larger binaries, particularly when limited to 1,000 transformations.
Based on these observations, we believe that a greater number of stacked transformations would further contribute to eventually eliminating the attacks associated with ret2spec and pht.

\begin{tcolorbox}[boxrule=1pt,arc=.3em,boxsep=-1.3mm]
  \textbf{Answer to \ref{rq:defensive}}:   Software diversification is effective at synthesizing \wasm binaries that mitigate Spectre-like attacks.  
  \tool generates variants of btb\_breakout and btb\_leakage that are protected against the considered attack. For ret2spec and pht, it generates hardened variants that are more resilient to the attack than the original program: 70\% of the diversified variants exhibit reduced attack effectiveness (reduced data leakage bandwidth) compared to the original program.
\end{tcolorbox}

\section{Discussion}
\label{discussion}

\textbf{Fuzzing \Wasm compilers with \tool}
In fuzzing campaigns, generating well-formed inputs is a significant challenge \cite{7958599}. 
This is particularly true for fuzzing compilers, where the inputs should be executable yet complex enough programs to probe various compiler components. 
\tool could address this challenge by generating semantically equivalent variants from an original \wasm binary, enhancing the scope and efficiency of the fuzzing process. 
A practical example of this occurred in 2021, when this approach led to the discovery of a wasmtime security CVE \cite{CVE}. 
Through the creation of semantically equivalent variants, the spill/reload component of Cranelift was stressed, resulting in the discovery of the before-mentioned CVE.

\textbf{Mitigating Port Contention with \tool:} 
Rokicki et al. \cite{10.1145/3488932.3517411} showed the practicality of a covert side-channel attack using port contention within \Wasm code in the browser. This attack fundamentally relies on the precise prediction of WebAssembly instructions that trigger port contention.
To combat this security concern, \tool could be conveniently implemented as a browser plugin. 
\tool can replace the \wasm instructions used as port contention predictor with other instructions.
This would inevitably remove the port contention in the specific port used to conduct the attack, hardening browsers against such attacks.

\begin{revision1}
    
    \textbf{Threats to validity:} We have observed several threats to validity related to \tool.
    First, the size of the variants, as demonstrated in the response to \ref{rq:performance}, is generally larger than the original. 
    This results in the potential for increased compilation time, cold spawn, and memory usage by the machine code. 
    Second, the rewriting rules of \tool are not easily extendable, except for manual addition and subsequent recompilation of the tool. 
    Third, the mutation backtracking of \tool does not account for the state of the mutation, i.e., \tool does not retain parsing information, the CFG, or the data flow graph for a previously observed binary. 
    This leads to the need to parse the same binary each time \tool conducts a transformation, which, as discussed in the response to \ref{rq:static}, slightly affects the number of generated variants. 
    Fourth, the preservation of variants, as indicated in \ref{rq:static}, is comparatively low concerning other diversification tools such as CROW \cite{arteaga2020crow}. 
    This is primarily because rewriting rules are manually crafted. 
    Unlike the transformations that CROW can generate, these can be more easily removed by compiler optimizations.
    Fifth, \tool and the tools upon which we built it may contain bugs. 
    For instance, the construction of the e-graphs heavily depends on the parsing of the Wasm binaries, which in turn relies on multiple other libraries. 
    However, to promote the auditing of our work and for the sake of open science, we have made the code freely available.

    
\end{revision1}

\section{Related Work}
\label{rw}


Static software diversification refers to the process of synthesizing, and distributing unique but functionally equivalent programs to end users. 
The implementation of this process can take place at any stage of software development and deployment - from the inception of source code, through the compilation phase, to the execution of the final binary \cite{jackson2011compiler, lundquist2016searching}.
\tool, a static diversifier, can be placed at the final stage, keeping in mind that the code will subsequently undergo final compilation by JIT compilers.
The concept of software diversification owes much to the pioneering work of Cohen \cite{cohen1993operating}. 
His suite of code transformations aimed to increase complexity and thereby enhance the difficulty of executing a successful attack against a broad user base \cite{cohen1993operating}. 
\tool's rewriting rules draw significantly from Cohen and Forrest's seminal contributions \cite{cohen1993operating, 595185}.

\begin{revision1}
As far as we know, Tigress is the only cutting-edge frontend diversifying virtualizer/obfuscator that supports \Wasm \cite{10.1145/3176258}.
Tigress's approach transforms the C code, maintaining it suitable for compilation using Emscripten, a C-to-Wasm compiler, resulting in a WebAssembly/html/Javascript package.
In general, applying diversification at the frontend has limitations. 
First, it would require a unique diversification mechanism for each language compatible with the frontend component.
Even though C/C++ code is the most frequently ported-to \Wasm language \cite{hilbig2021empirical}, our needs involve modifying any Wasm in existence.
Second, source-based diversification tends to alter the code section of the final \Wasm binary more significantly.
As a result, other sections of the \Wasm binaries receive less attention, or even remain untouched, during the diversification process.
Yet, \tool, can modify any section of any \Wasm binary in existence.
Third, source code diversification could unintentionally introduce compiler fingerprints into the compiled \Wasm \cite{CABRERAARTEAGA2023103296}.

\end{revision1}

Jackson and colleagues proposed the pivotal role of the compiler in promoting static software diversification \cite{jackson2011compiler}.
Within the \wasm context, CROW is the only existing compiler-based diversifier \cite{arteaga2020crow}.
It is recognized as a superdiversifier for \wasm, built within the LLVM compilation toolchain \cite{jacob2008superdiversifier}.
However, the direct integration of the diversifier into the LLVM compiler restricts its applicability to \wasm binaries generated through LLVM.
This limitation implies that \wasm source code without an LLVM frontend implementation cannot leverage CROW's capabilities.
Conversely, \tool provides a more adaptable and expedited \wasm to \wasm diversification solution, ensuring compatibility with any compiler.
Additionally, unlike CROW, \tool does not depend on an SMT solver to validate the generated variants.
It instead guarantees semantic equivalence by design, leading to increased efficiency in generating \wasm variants, as discussed in \autoref{rq:static:results}.
\revision{
Consequently, CROW seems to provide variants that are more resilient to further compiler optimizations.
\tool trades off the preservation of generated variants to create more variants.
According to CROW reports, \tool generates more variants by at least one order of magnitude in the same amount of time. 
}
Overall, as a \wasm to \wasm diversification tool, \tool extends the range of tools capable of generating \wasm programs, a topic thoroughly explored in this work.

The process of diversifying a \Wasm program can be conceptualized as a three-stage procedure: parsing the program, transforming it, and finally re-encoding it back into \wasm. 
Our review of the literature has revealed several studies that have employed parsing and encoding components for \wasm binaries across various domains. 
This indicates that these works accept a \wasm binary as an input and output a unique \wasm binary. 
These domains span optimization \cite{wasmslim}, control flow \cite{10123627}, and dynamic analysis \cite{wasabi, stievenart2020compositional, 10123627, BRITO2022102745}.
When the transformation stage introduces randomized mutations to the original program, the aforementioned tools could potentially be construed as diversifiers.
\tool is related to these previous works, as it can serve as an optimizer or a test case reducer due to the incorporation of an e-graph at the heart of its diversification process \cite{10.1145/1480881.1480915}. 
To the best of our knowledge, the introduction of an e-graph into \tool marks the first endeavor to integrate an e-graph into a \wasm to \wasm analysis tool.


BREWasm \cite{rewritingtool2023} offers a comprehensive static binary rewriting framework for \Wasm and can be considered to be the most similar to \tool. 
For instance, it can be used to model a diversification engine.
It parses a WebAssembly binary into objects, rewrites them using fine-grained APIs, integrates these APIs to provide high-level ones, and re-encodes the updated objects back into a valid WebAssembly binary. 
The effectiveness and efficiency of BREWasm have been demonstrated through various WebAssembly applications and case studies on code obfuscation, software testing, program repair, and software optimization. 
The implementation of BREWasm follows a completely different technical approach.
In comparison with our work, the authors pointed out that our tool employs lazy parsing of WebAssembly. 
Although they perceived this as a limitation, it is eagerly implemented to accelerate the generation of \wasm binaries.
Additionally, our tool leverages the parser and encoder of wasmtime, a standalone compiler and interpreter for WebAssembly, thereby boosting its reliability and lowering its error-prone nature.

Another similar work to \tool is WASMixer \cite{wasmixer}.
WASMixer focuses on three code obfuscation methods for WebAssembly binaries: memory access encryption, control flow flattening, and the insertion of opaque predicates. 
Their strategy is specifically designed for obfuscating WebAssembly binaries. 
In contrast, while \tool does not employ memory access encryption or control flow flattening, it can still function effectively as an obfuscator. 
Previous evaluations confirm that \tool has been successful in evading malware detection \cite{CABRERAARTEAGA2023103296}.
On the same topic, Madvex \cite{madvex} also aims to modify WebAssembly binaries to achieve malware evasion, but their approach is principally driven by a generic reward function and is largely confined to altering only the code section of a WebAssembly binary. 
\tool, however, adopts a more flexible strategy by applying a broader array of transformations, which are not limited to the code section. 
Consequently, \tool is capable of generating malware variants without negatively affecting their code or performance.

\section{Conclusion}
\label{conc}

\tool is a fast and effective diversification tool for \wasm, with a 100\% diversification rate across the 303 programs of the considered benchmark. 
Concerning speed, it creates over 9000 unique variants per hour.
The \tool workflow ensures that all final variants offer different and unique execution traces. 
\revision{Remarkably, \tool creates variants that have a minimal impact on execution time.}
We have proven that \tool can mitigate Spectre attacks in \wasm, producing fully protected variants of two versions of the btb attack, and variants of ret2spec and pht that leak less data than the original ones.

In future work, we aim to fine-tune the diversification process, balancing broad diversification with the needs of specific scenarios. 
Besides, the creation of rewriting rules for \tool is currently a manual task, yet we have identified potential for automation. 
For instance, \tool could be enhanced through data-driven methods such as rule mining.
Furthermore, we have observed that the impact of \tool on ret2spec and pht attacks is considerably less compared to btb attacks. 
These attacks exploit the returning address of executed functions in the program stack. 
One mitigation of this would be multivariant execution strategy, implemented on top of \tool. By offering different execution paths, the returning addresses on the stack at each function execution would vary, thereby improving the hardening of binaries against ret2spec attacks.

\bibliographystyle{cas-model2-names}

\bibliography{main}

\balance

\end{document}